\documentclass[12pt,preprint]{aastex}
\usepackage{natbib,graphicx}

\slugcomment{To appear in ApJ}

\shorttitle{30 AU radius CO gas hole around IRS 48}
\shortauthors{Brown et al.}

\begin{document}


\title{A 30 AU radius CO gas hole in the disk around the Herbig Ae star Oph IRS 48\footnote{This work is based on observations collected at the European Southern Observatory Very Large Telescope under program ID 179.C-0151.}}


\author{{J.M. Brown\altaffilmark{1,2}, G.J. Herczeg\altaffilmark{1}, K.M. Pontoppidan \altaffilmark{3}, E.F. van Dishoeck\altaffilmark{1,4}
}}
\altaffiltext{1}{Max-Planck-Institut f{\"u}r extraterrestrische Physik, Postfach 1312,  85741 Garching, Germany ; joannabrown@cfa.harvard.edu}
\altaffiltext{2}{Harvard-Smithsonian Center for Astrophysics, 60 Garden St., MS 78, Cambridge, MA 02138}
\altaffiltext{3}{Space Telescope Science Institute, 3700 San Martin Drive, Baltimore, MD 21218}
\altaffiltext{4}{Leiden Observatory, P.O. Box 9513, NL-2300 RA Leiden, The Netherlands}



\begin{abstract} 
The physical processes leading to the disappearance of disks around
young stars are not well understood. A subclass of transitional disks,
the so-called cold disks with large inner dust holes, provide a
crucial laboratory for studying disk dissipation processes. IRS 48 has
a 30 AU radius hole previously measured from dust continuum imaging at
18.7 $\mu$m. Using new optical spectra, we determine that IRS 48 is a
pre-main sequence A0 star. In order to characterize this disk's gas
distribution, we obtained AO-assisted VLT CRIRES high resolution (R
$\sim$ 100,000) spectra of the CO fundamental rovibrational band at
4.7 micron. All CO emission, including that from isotopologues and
vibrationally excited molecules, is off-source and peaks at 30 AU. The
gas is thermally excited to a rotational temperature of 260 K and is
also strongly UV pumped, showing a vibrational excitation temperature
of $\sim$5000 K. We model the kinematics and excitation of the gas and
posit that the CO emission arises from the dust
hole wall. Prior imaging of UV-excited PAH molecules, usually a
gas tracer, within the hole makes the large CO hole even more
unexpected.

\end{abstract}




\section{Introduction}

Disk around pre-main sequence stars are the birthsites of planets, yet
many questions remain about the formation processes of planets and the
end stages of the disk. Planets are expected to leave distinctive gaps
in the disk as they clear their orbital paths \citep{artymowicz94,
bryden99, crida07}. Transitional disks, in which dust clearing has
begun in the inner regions, may be the predicted objects
(e.g. \citealt{strom89, calvet02, brown07, najita07}). A young
substellar candidate recently discovered in the transitional disk T
Cha lends credence to this hypothesis (\citealt{huelamo11}). However,
dust deficits in spectral energy distributions (SEDs) can potentially
result from other processes including photoevaporation and grain
growth. Gas observations can help to distinguish between these
scenarios. Gas and dust are removed near simultaneously in inside-out
disk photoevaporation models and would have similar distributions
\citep{alexander06,owen11}. At the other extreme, the creation of an
apparent dust hole through grain growth and settling would cause no
depletion of the gas. Finally, stellar and planetary companions can
clear gaps free of both gas and dust, but some movement of material
across these gaps would be expected depending on companion size and
location \citep{artymowicz96,zhu11}. The creation of gaps by
companions can enhance other avenues of disk clearing such as
photoevaporation \citep{alexander09}. Gas observations are also of
particular relevance in the case of young planets as the gas drives
both formation and migration.

CO rovibrational emission is sensitive to warm $\sim$100-1000 K gas
and is commonly seen from young disks around both Herbig Ae/Be
(e.g. \citealt{blake04}) and T Tauri stars (\citealt{najita03}). These
lines are particularly useful as tracer of gas for transition disks as
the $\sim$1000 K gas temperatures and high velocities likely imply gas
within the dust holes \citep{salyk09}. Spectroastrometric observations
of three transition disks confirm that the CO fundamental emission
generally traces gas within the dust holes but in one case a 7 AU gas
hole within a $\sim$30 AU dust hole was found
\citep{pontoppidan08}. Lack of CO fundamental emission from the inner
regions of Herbig Ae/Be stars appears slightly more common with a
handful of disks currently known: HD 141569A (11 AU radius gas hole,
\citealt{goto06}), HD 97084 and HD 100546 (11 and $>$8 AU radii
respectively, \citealt{vanderplas09}). \citet{brittain07,brittain09}
successfully modeled the emission distributions in both HD 141569A and
HD 100546 with UV-fluorescent gas. \citet{pontoppidan11b} finds that
it is natural for higher luminosity sources to have larger inner
CO-free regions, although some sources have even larger inner holes
than expected.

The radiative environment around young high mass stars is harsher than
around T Tauri stars and allows investigation of disk dissipation
processes under different conditions. The majority of transitional
disks found to date surround G-type and later stars, although dust
holes are seen at all stellar masses. Many factors might be responsible for the lower numbers
including the rarity of local Herbig Ae/Be stars, the potential for a
faster transition timescale, and difficulty in identifying transition
disks around higher mass stars. Herbig Ae/Be SEDs have traditionally
been classified into almost flat SEDs (group I) and those with a
strong decline in the far-infrared (group II) \citep{meeus01,
hillenbrand92}. These classes represent flared and self-shadowed disks
\citep{dullemond04}, although the transition between the two is likely
continuous \citep{meijer08}. The high luminosities of these stars are
capable of producing puffed-up inner rims which shadow the disk behind
\citep{dullemond04, acke09} and can produce SEDs qualitatively similar
to transition disk SEDs due to the difficulty in distinguishing
between shadowed and missing material. However, dust holes are present
in disks around at least some higher mass sources with millimeter
imaging of MWC 758 (A8) and HD 135344B (F4) clearly resolving dust
cavities \citep{andrews11,brown09}. Exoplanets have also been directly
imaged around older high mass stars at distances that could explain large
inner holes (e.g. \citealt{marois08}) implying that their younger precursors may be transition disks. 

The Herbig Ae star IRS 48 in $\rho$ Ophiuchi star formation region
has an unusual disk. It was first noted for its strong PAH emission
and implied strong UV field. Spatially resolved mid-IR images of the
dust reveal a ring-like structure at 18.7 \micron\, with a
distance of 110 AU between the brightness peaks and an inner dust hole
with a radius of $\sim$30 AU \citep{geers07}. A strong east-west
directionality is seen leading \citet{geers07} to conclude a disk
inclination of 48$^{\circ}$$\pm$8$^\circ$. Strong polycyclic aromatic
hydrocarbon (PAH) molecular emission, likely tracing gaseous molecular
material, was found to be centered on the star and fill in this large
grain ring. The CO gas might be expected to have a similar
distribution to the PAHs and be present inside the dust hole. The
VLT-CRIRES spectrograph offers the combination of spectral and spatial
resolution to resolve the warm gas in this disk. In this paper, we
present CRIRES observations of IRS 48 showing spatially extended CO
rovibration emission and examine both the kinematics and excitation to
determine the gas structure and origin.

\section{Observations}

\subsection{CRIRES} 

We observed Oph IRS 48 (WLY 2-48, $\alpha$ = 16$^{\rm h}$27$^{\rm m}$37$^{\rm s}$.19, $\delta$ =
-24$^{\rm \circ}$30$^{\rm '}$34\farcs 8, J2000) with CRIRES in the M band (4.63-4.87 \micron)
covering the CO fundamental rovibrational band as part of a large
CRIRES program studying a sample of $\sim$ 100 stars
\citep{pontoppidan11a}. CRIRES is a cryogenic echelle spectrograph
behind the Multi-Application Curvature Adaptive Optics (MACAO) system
on the 8m VLT 1, Antu \citep{kaeufl04}. A slit width of 0\farcs2 was
used to provide a resolving power of R $\sim$ 100,000 ($\Delta$ v = 3
km s$^{-1}$). The observations were taken on several different dates
in 2007 and 2008 (see Table \ref{table:obs}).

The data were reduced following the procedures of
\citet{pontoppidan08}. The data were taken with an ABBA 10" dithered
nodding pattern permitting a first order correction of the infrared
background by pair subtraction. Integration times for each position
were 1 minute leading to 4 minute nod cycles. The number of nod cycles
in each setting is listed in Table \ref{table:obs}. The CRIRES pixel
scale was 86 mas which was subdivided during reduction into 43
mas/pixel.  The 2-D spectra were linearized and then extracted. The
spatially extended emission required customized extraction. For an
overall spectrum, an aperture of 1\farcs2 was used to include all
extended emission. Spectra were also extracted in 2 pixel (86 mas)
apertures to produce a spectral map. All spectra were wavelength
calibrated using the strong telluric features in the standard star
spectra. The standard stars all have spectral types of O or B, chosen
for their lack of spectral features in the M band. Finally, the many
weak telluric features were removed by dividing the source spectra by
the standard star spectra. Remaining strong telluric features with
transmission of $<$20\% were blanked from the analyzed spectra.

The majority of the observations were taken with the slit at a
position angle of 90$^\circ$, following the east-west directionality
seen in the \citet{geers07} images. Spectra at 0$^\circ$ and
180$^\circ$ were also taken to examine the orthogonal distribution of the CO
emission (see Table \ref{table:obs}).

Flux calibration was done by normalizing to Spitzer IRAC
fluxes. Calibration directly from the M-band standard star spectra
suffers from variable throughput effects, but in this case, direct
flux calibration was within 20\% of the Spitzer IRAC fluxes. We
therefore have reasonable confidence that the 5 $\mu$m continuum flux
has not changed significantly since the Spitzer observations.

\subsection{Spectral type of IRS 48}
\label{stellarparams}

The spectral type of IRS 48 is unclear from the
literature. \citet{luhman99} found a spectral type earlier than F3
from a K-band spectrum.  In contrast, \citet{geers07} argued for M0
based on an optical spectrum obtained with a fiber-fed multi-object
spectrograph.  However, telluric H$_2$O absorption in the optical
spectrum may have been mistaken for the TiO band absorption
characteristic of M0 stars.

\begin{figure}[h!]
\begin{center}
\includegraphics[angle=0,scale=0.7]{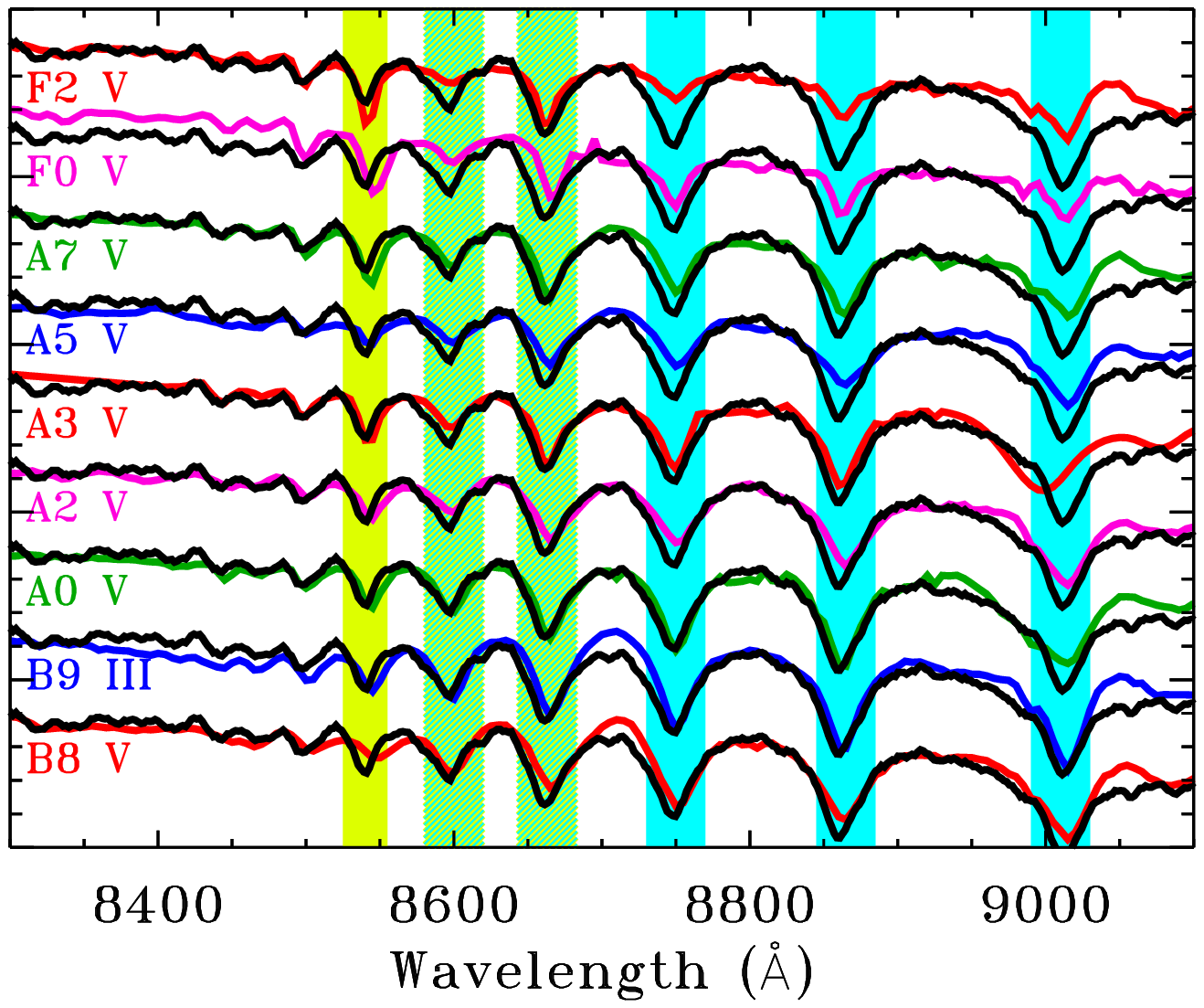} 
\end{center}
\caption{Comparison of the H Paschen series in the IRS 48 LRIS-ADC
spectrum (black) with the \citet{pickles98} spectral library. The
presence and strength of these lines clearly indicates that IRS 48 is
an A-type star, with a best fit of A0$^{+4}_{-1}$. \label{fig:sptype}}
\end{figure}

To resolve this discrepancy, we obtained a flux-calibrated,
low-resolution ($R\sim1000$) optical spectrum using LRIS-ADC on Keck I
\citep{oke95} on 2008 May 28.  Although the observation spans
3000--9300 \AA, the star is only detected longward of 5500 \AA\
because of extinction.  Some weak H$\alpha$ emission is tentatively
detected in our spectrum and in the \citet{geers07} spectrum, likely
indicating the presence of accretion and consequently gas in the inner
disk \citep{muzerolle04}. Pfund-$\beta$ emission is also detected in
the CRIRES spectra, another probable indicator of accretion. The
presence and relative strength of absorption in the H Paschen and
\ion{Ca}{2} IR triplet lines clearly indicate that IRS 48 is an A-type
star, consistent with the Br-$\gamma$ absorption detected by
\citet{luhman99}.  By comparison with the \citet{pickles98} stellar
spectral flux library, we assign a spectral type of A0$^{+4}_{-1}$ to
IRS 48, which corresponds to an effective temperature of 9000$\pm$550
K (Figure \ref{fig:sptype}).

The extinction is estimated by comparing the continuum slope in our
optical spectrum to the continuum slopes of early A-type stars in
\citet{pickles98}.  The extinction curve used is that of
\citet{weingartner01} for $R_V=5.5$, which approximates the opacity of
large dust grains that are thought to be typical of dense star-forming regions
\citep{indebetouw05}.  We measure $A_V=11.50\pm0.25$ mag, confirming
the presence of substantial intervening material in the line of sight,
as is common in Ophiuchus.

Based on the spectral type, extinction, a J-band magnitude of 10.57
\citep{skrutskie06}, and a distance of 120 pc to $\rho$ Oph
\citep{loinard08}, the stellar luminosity is 14.3 $L_\odot$. The
\citet{siess00} evolutionary tracks then imply that IRS 48 is a $\sim
2.0$ M$_\odot$ star with an age of $\sim 15$ Myr at the start of the
main sequence.  The spectral type and luminosity are inconsistent with
the 1 Myr age that is typically measured for Ophiuchus.  At 1 Myr, a
star with spectral type A0 would have a luminosity of $\sim 200$
$L_\odot$ and a mass of $\sim 4.5 M_\odot$. Potential sources of error
in the calculated luminosity could be a high gray extinction and
scattering off an edge-on disk, the dust opacity ($R_V$=5.5) used in
the extinction correction or the distance, which would likely exclude
membership in Ophiuchus. We discuss this discrepancy further in
Section \ref{discussion}, including additional constraints from the
CRIRES data.

\section{Analysis}
\label{analysis}

\begin{figure*}[h!] 
\hspace{-2cm} 
\begin{minipage}{0.11\linewidth}
\includegraphics[angle=90,scale=0.3]{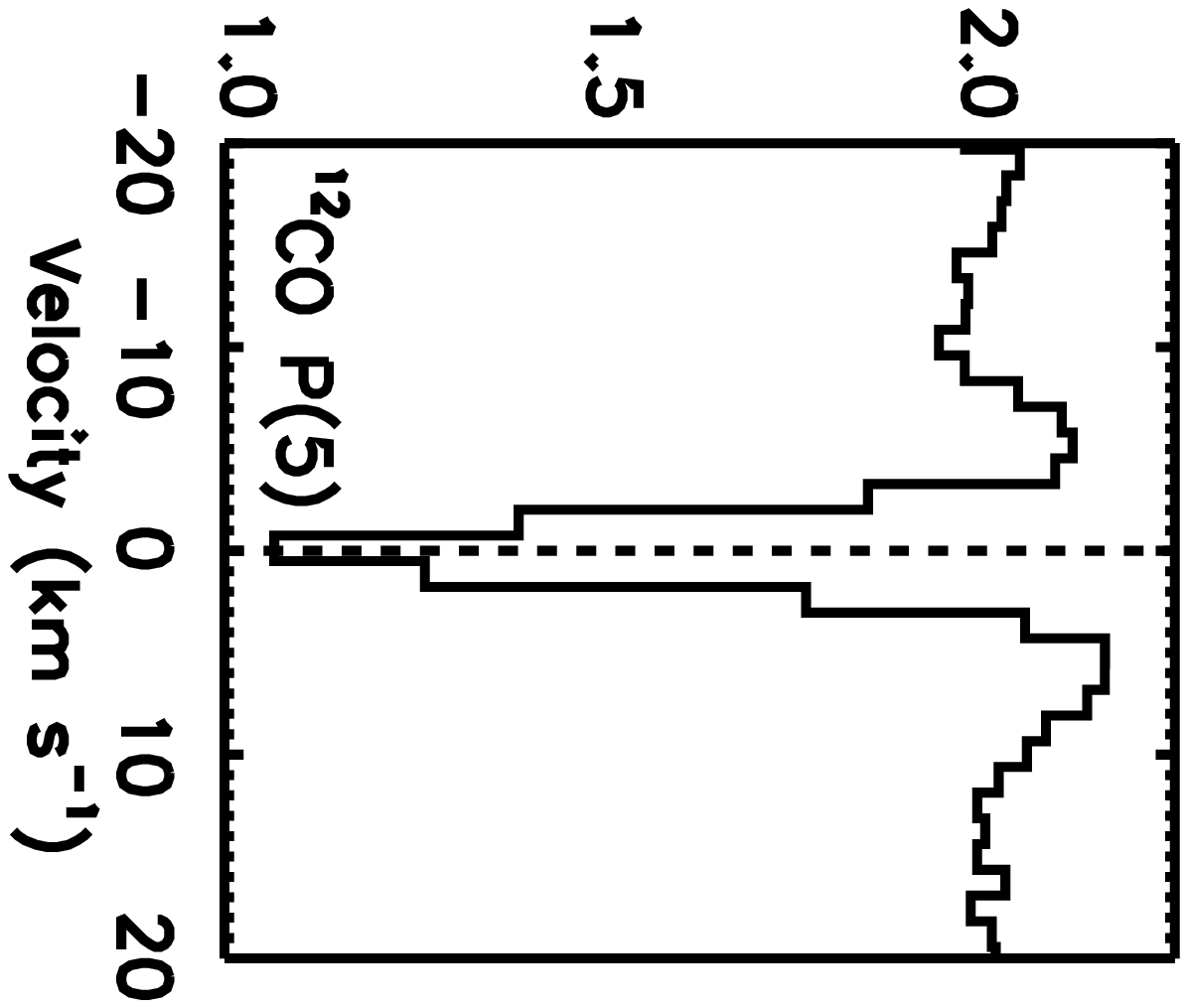} 
\end{minipage}
\begin{minipage}{0.11\linewidth}
\includegraphics[angle=90,scale=0.3]{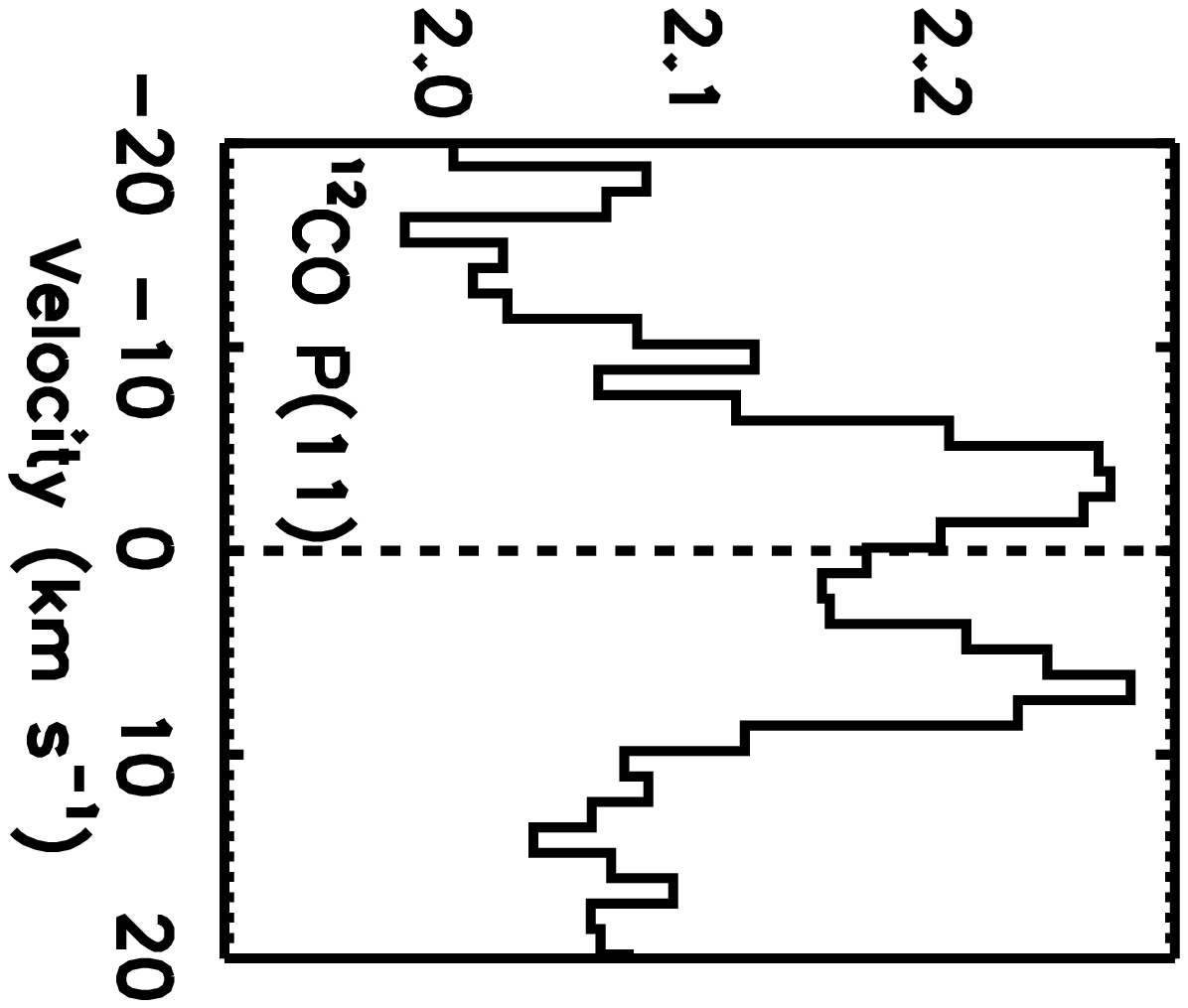} 
\end{minipage}
\begin{minipage}{0.11\linewidth}
\includegraphics[angle=90,scale=0.3]{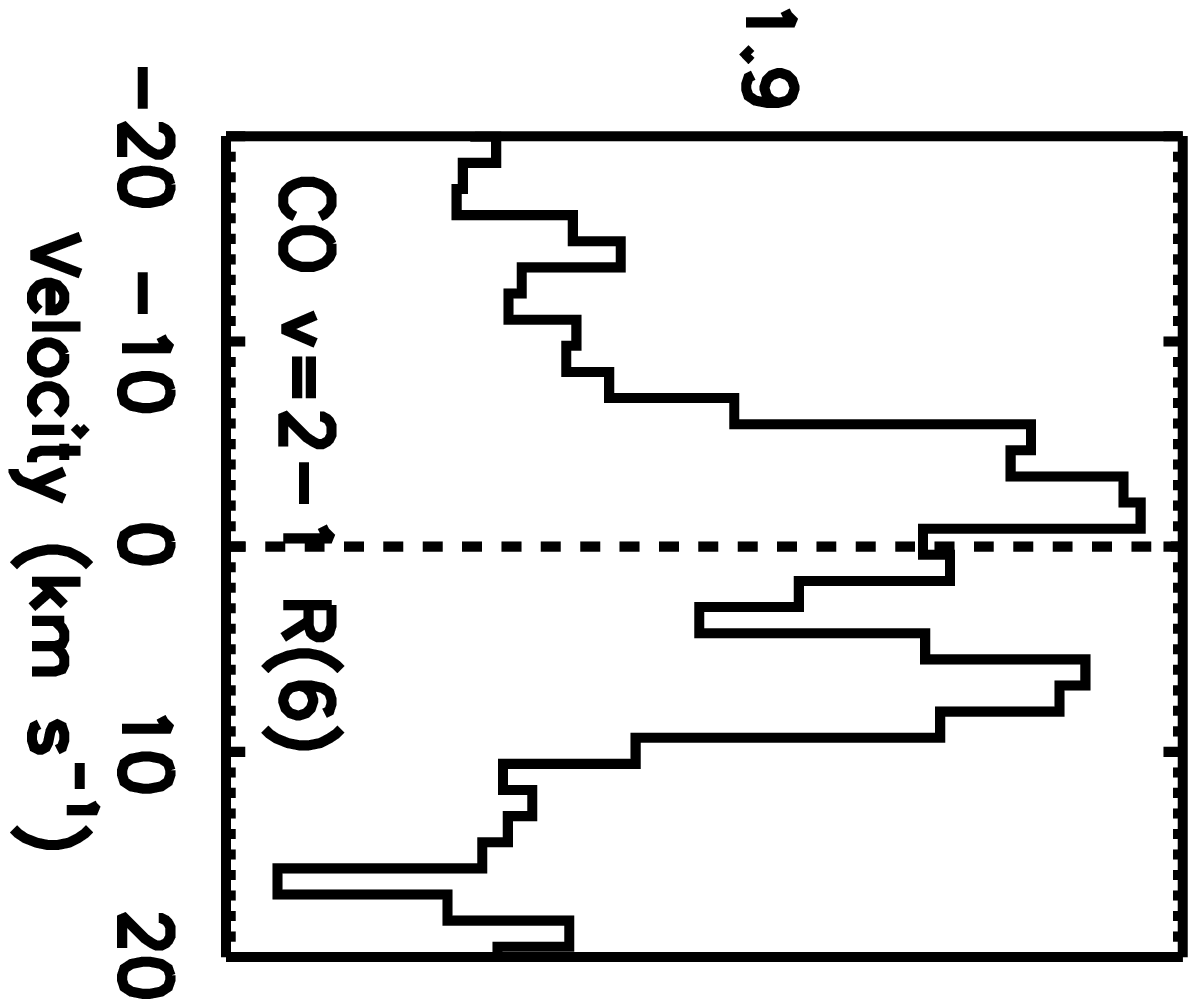} 
\end{minipage}
\begin{minipage}{0.11\linewidth}
\includegraphics[angle=90,scale=0.3]{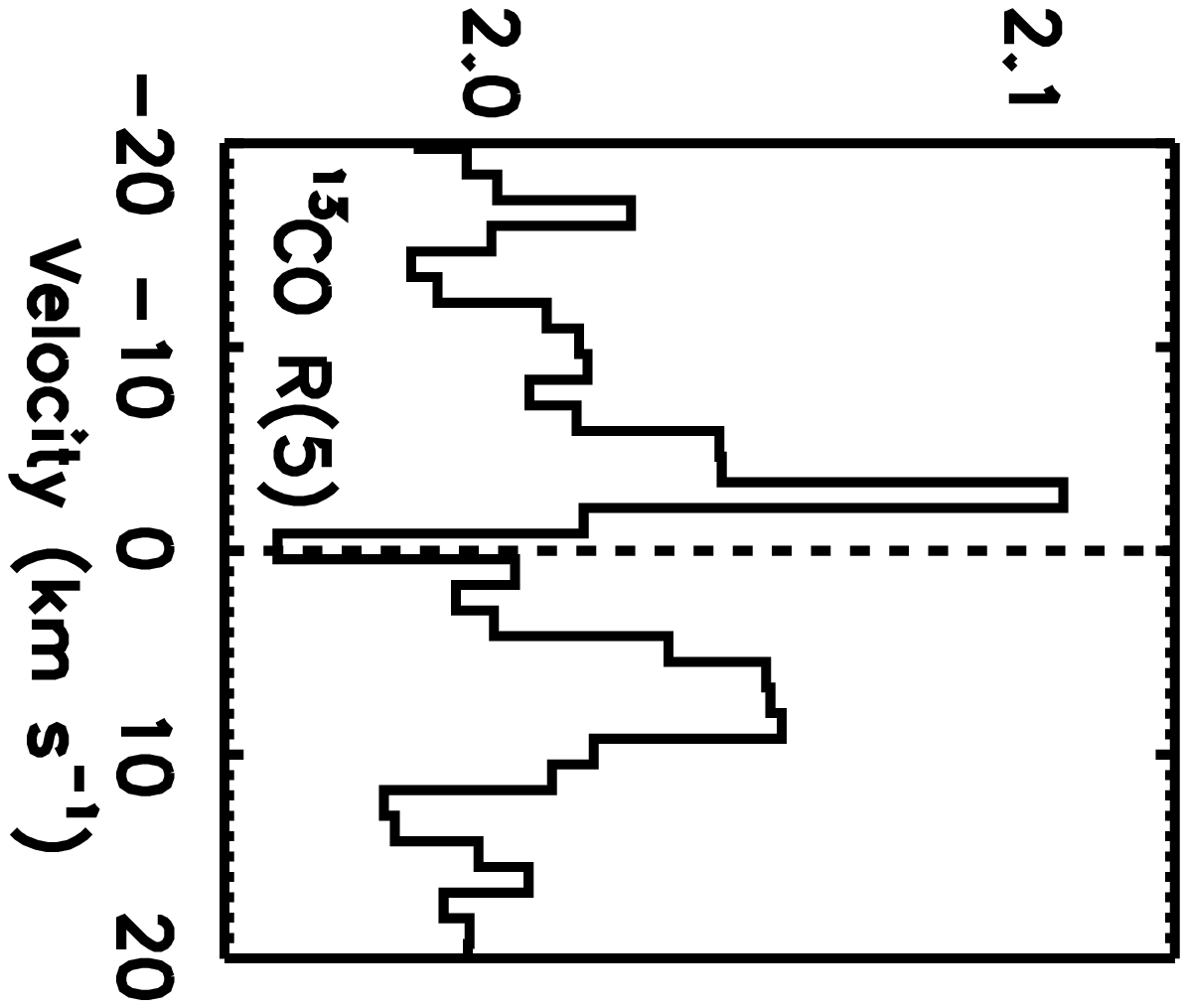} 
\end{minipage}
\begin{minipage}{0.25\linewidth}
\includegraphics[angle=90,scale=0.3]{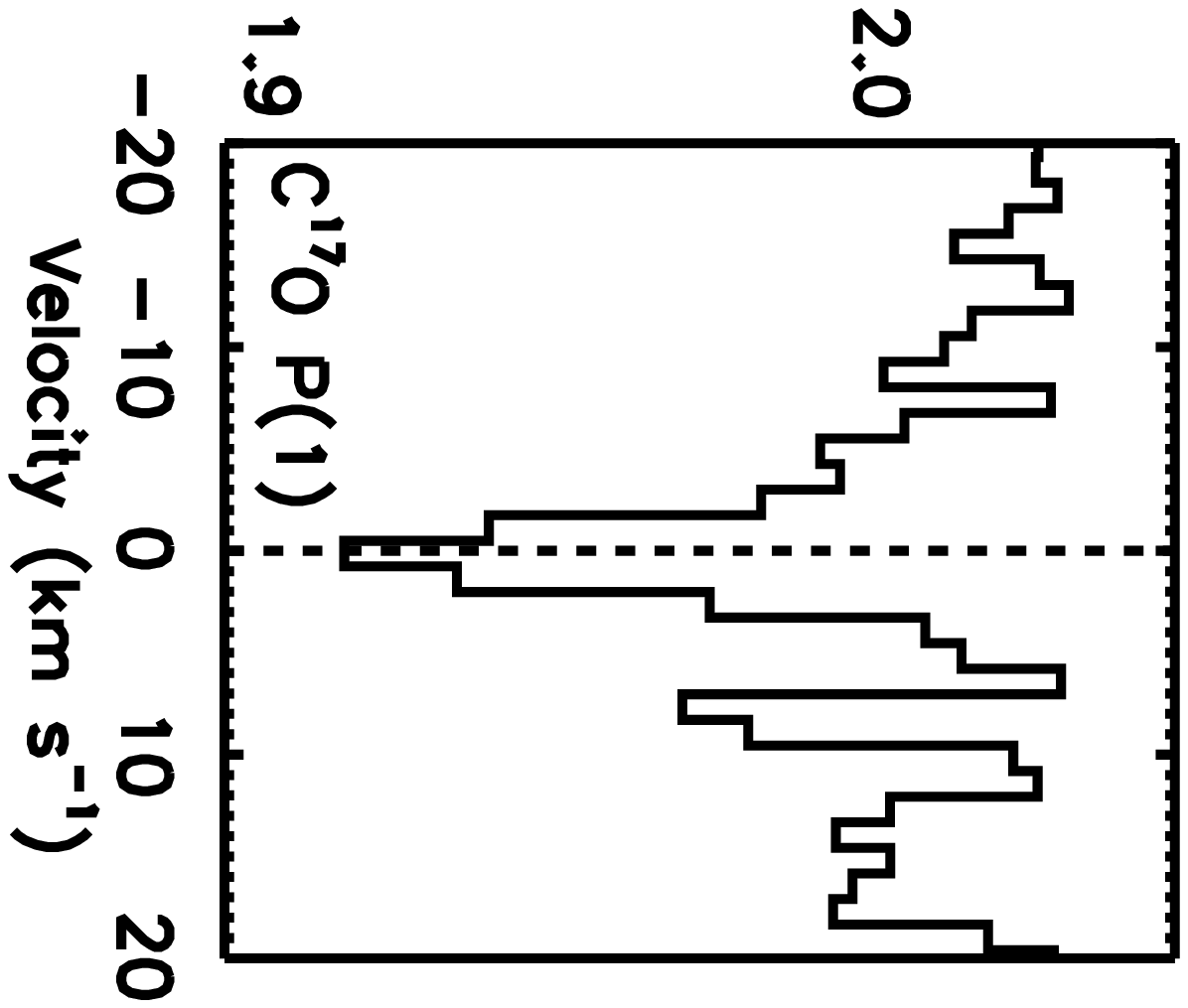} 
\end{minipage}
\vspace{0.01cm} 
\begin{minipage}{1.\linewidth}
\hspace{-1cm}
\includegraphics[angle=90,scale=0.65]{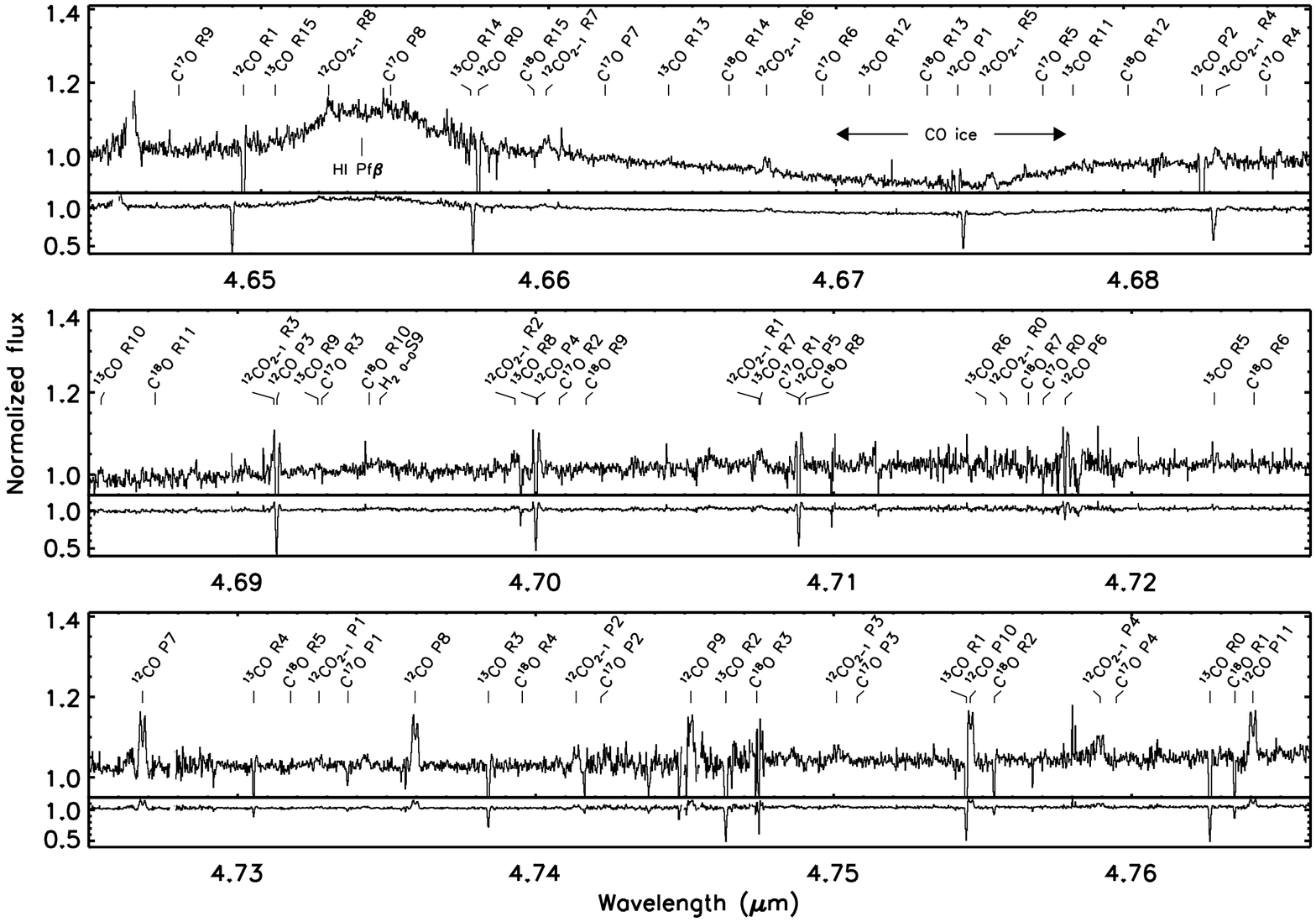} 
\end{minipage}
\caption{VLT-CRIRES spectrum of IRS 48 with the upper row showing
close ups of different lines and the lower panels showing the emission
lines in detail on the top and the full depth of the absorption lines
on the bottom. $^{12}$CO,$^{13}$CO and C$^{18}$O are seen in
absorption, particularly strong in low energy $J$ transitions. The
$^{12}$CO v=1-0 and v=2-1 lines show strong double peaked emission
features. \label{fig:spectrum}} \end{figure*}

Both CO emission and absorption are detected from IRS 48 (Figure
\ref{fig:spectrum}). CO emission is seen in the v=2-1 and very weak
v=3-2 lines, while the isotopologues, C$^{18}$O and C$^{17}$O, are in
absorption. The $^{12}$CO and $^{13}$CO v=1-0 lines contain a blend of
the two components with absorption at the line center and emission at
higher velocities. The absorption lines disappear at energy levels
higher than P(6), indicating that the absorbing gas is cold with an
excitation temperature of $<$30 K. The emission lines are all well
resolved spectrally with CRIRES and are double peaked, including more
excited transitions free from absorption. Weak CO ice absorption
produces a broad shallow feature between 4.66 and 4.685 $\mu$m and the
HI Pf$\beta$ line is observed at 4.653 $\mu$m, underlying the CO v=2-1
R(8) line. The H$_2$ v=0 S(9) line at 4.6947 $\mu$m is not detected.

\begin{figure*}[h!]
\begin{minipage}{0.6\linewidth}
\vspace{-0.5cm}
\includegraphics[angle=0,scale=0.45]{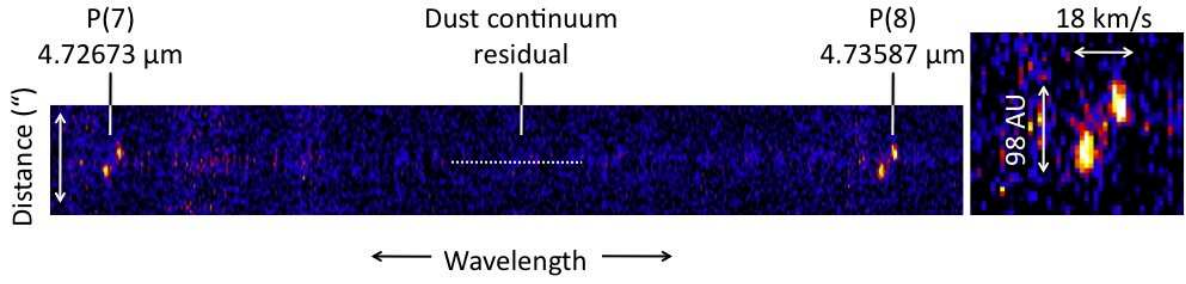}
\end{minipage}
\vspace{-8.cm}
\caption{Continuum subtracted 2D spectra showing the extended line
emission in the $^{12}$CO P(7) and P(8) lines from 2007 September
5. The emission peaks around 30 AU and is much weaker towards the
central position.  \label{fig:2d}}
\end{figure*}

The CRIRES AO system produes a small enough PSF in the spatial
direction that bright extended emission can be seen directly in
the sky-subtracted 2-D images. The continuum was subtracted based on
the mean continuum spatial profile and a scaled telluric standard was
divided out at every spatial position. The IRS 48 emission lines are
spatially extended in the 2-D spectra before extraction, particularly
clearly in the observations taken at a position angle of
90$^{\circ}$. Figure \ref{fig:2d} shows the $^{12}$CO extended
emission in the continuum subtracted 2-D spectral trace. 

\begin{figure*}[h!]
\begin{center}
\includegraphics[angle=90,scale=0.3]{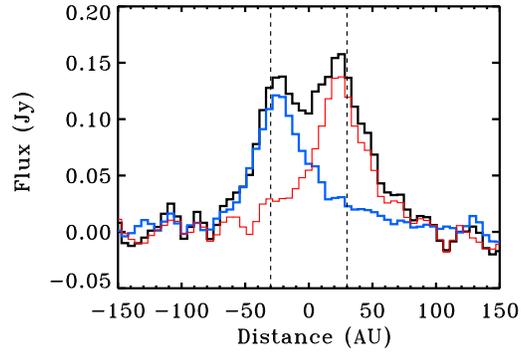}
\end{center}
\caption{Cross section of the stacked lines at PA=90$^\circ$. The
black line is the total spatial distribution while the blue and the
red lines are the blue and red sides of the line. The peak emission
averaged over the entire line width occurs at $\pm \sim$25 AU. This is
slightly smaller than the largest peak to peak extents, marked with
dashed vertical lines, due to the inclusion of gas not at apoapsis of
the inclined ring. When compared directly to the continuum
distribution, the CO spatial profiles are consistent with a small
emitting region (see also \S 4.2). \label{fig:spat}}
\end{figure*}

Extended emission is unusual among the larger sample of mostly T Tauri
stars studied with CRIRES, where most sources show no detectable
extended emission in the 2-D images (\citealt{brown11}, in prep.). The
lines are narrow with a width at line base of 18 km s$^{-1}$ while
most of the parent sample show line widths well over 50 km s$^{-1}$
unless the disk is known to be face-on. The narrow width, and thus
lack of high velocity gas, is consistent with gas motions from large
radii and limits the inner extent of the gas, assuming Keplerian
motion. Figure \ref{fig:spat} shows the spatial cross section of the
clean $^{12}$CO stacked lines.

\begin{figure}[h!]
\begin{minipage}{0.28\linewidth}
\includegraphics[angle=90,scale=0.6]{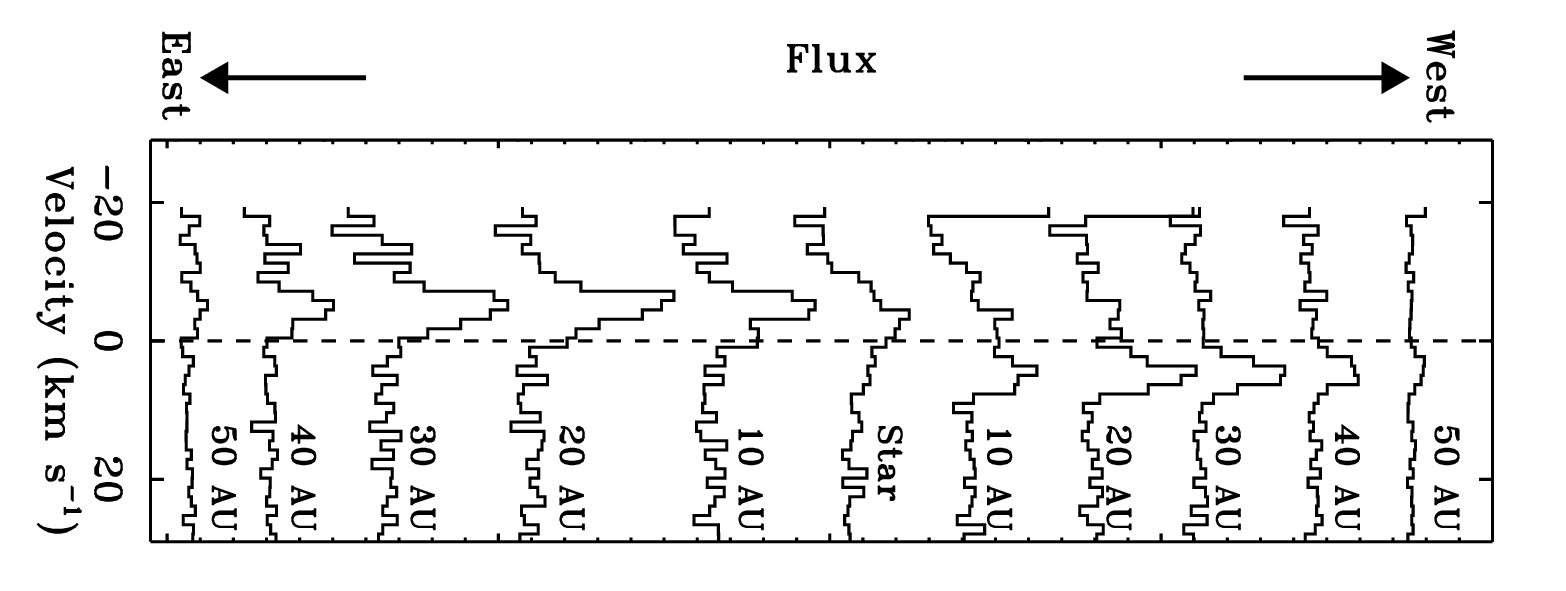} 
\end{minipage}
\begin{minipage}{0.28\linewidth}
\includegraphics[angle=90,scale=0.6]{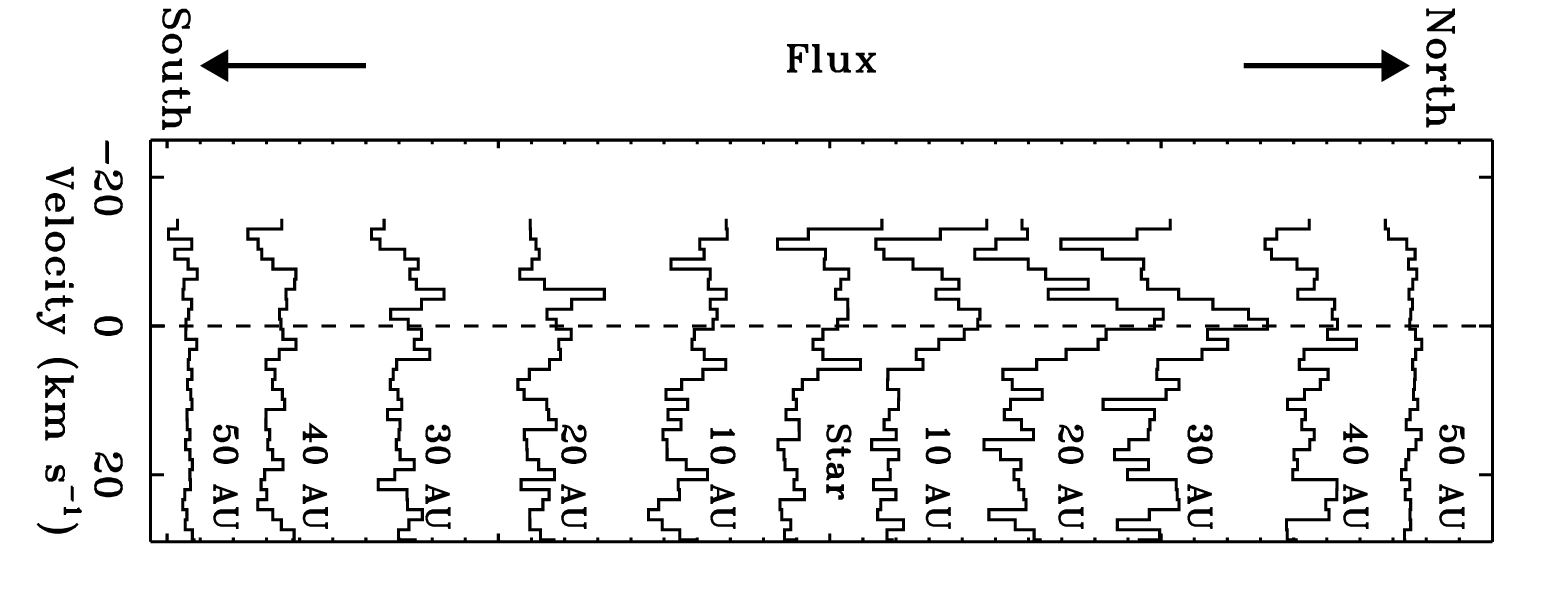} 
\end{minipage}
\begin{minipage}{0.45\linewidth}
\includegraphics[angle=0,scale=0.45]{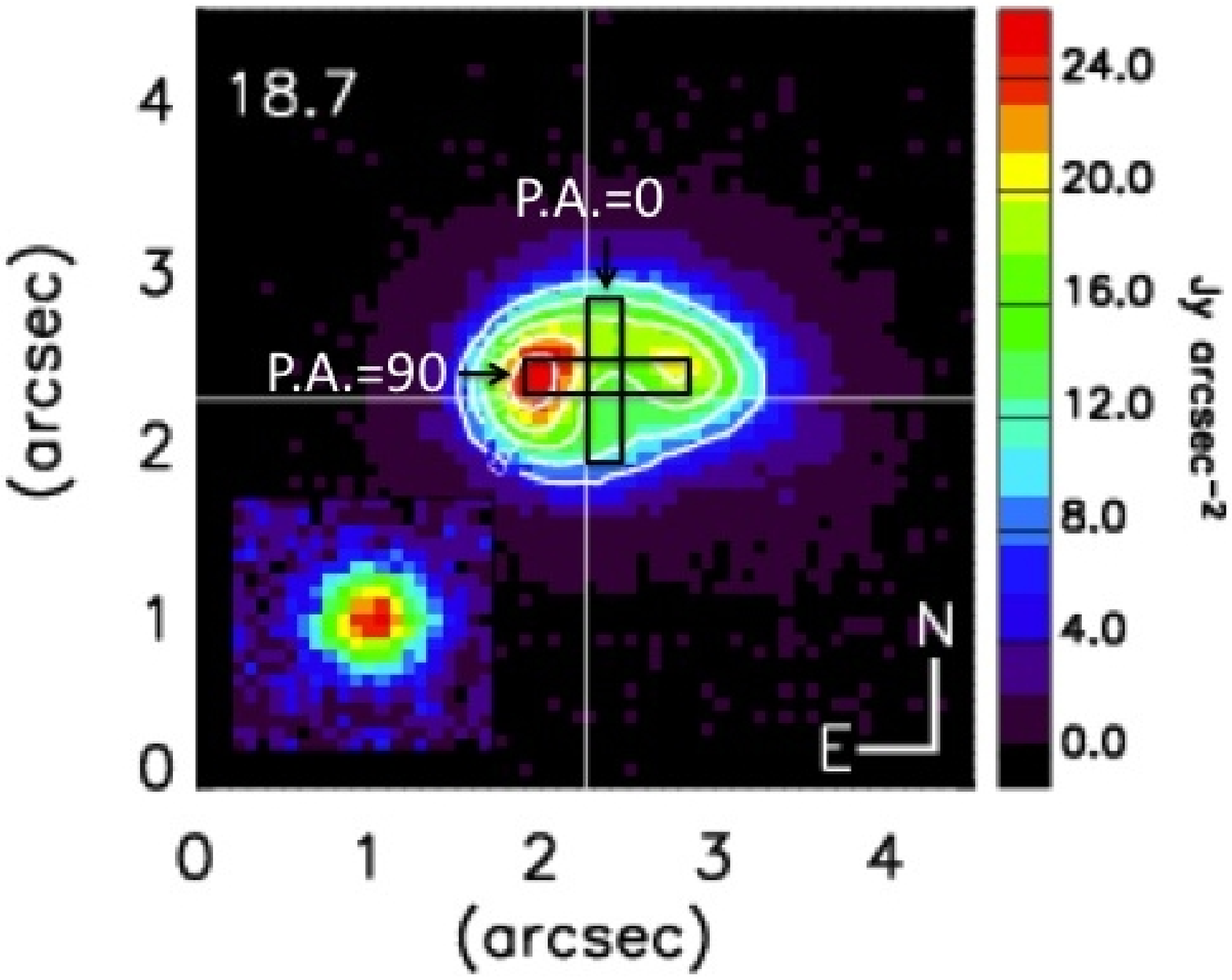} 
\end{minipage}
\caption{(Left) Spatial changes in the profile of the P(8) transition
of $^{12}$CO v=1-0 in increments of 86 mas (10 AU) for PA=90$^\circ$
(left) and PA=0$^\circ$ (right). Emission occurs from the outer
regions while line of sight absorption occurs towards the central star
at lower J ($<$6). Keplerian rotation can clearly be seen in the
emission with the emission peak in the top spectra redshifted compared
to the bottom. The missing data on the left, removed due to
contamination by the telluric P(8) line, show the clear separation
between science and sky lines. (Right) The slit positions overlaid on
the \citet{geers07} 18.7 $\mu$m image. Some of the asymmetries of the
dust are echoed in the gas, including brighter emission in
the eastern peak compared to the west and brighter emission from the
north compared to the south. \label{fig:p5}}
\end{figure}

By extracting the spectra on a per spatial pixel basis, the physical
location of the different components is probed (see Figure
\ref{fig:p5}). The CO lines on the central stellar position appear in
absorption in lines with J $<$ 7 and have little emission in all
transitions.  Strong $^{12}$CO v=1-0 and v=2-1 emission lines can be
seen at distances further from the star. Two components are visible: a
redshifted component to the west and a blueshifted component to the
east. The maximum flux is seen at $\sim$30 AU (6 pixels) from the
central star. The spectra taken at a position angle of 0$^\circ$ show
emission in both directions centered around 0 km s$^{-1}$ (Figure
\ref{fig:p5}), in agreement with the 90$^\circ$ position angle of the
ring seen in the 18.7 $\mu$m dust image
\citep{geers07}. There are, however, some asymmetries apparent in the
PA=0$^\circ$ data with the north side being brighter and more strongly
peaked towards the line center. A similar brightness asymmetry between
the north and south sides of the ring is seen in the 18.7 $\mu$m dust
 (Figure \ref{fig:p5}, \citealt{geers07}).

One additional calculation that can be made from the 2-D data is an
inclination estimate based on the assumption of a symmetrical tilted
disk. The major axis is determined to be 56 AU $\pm$ 2 AU peak to peak
and 120 AU $\pm$ 5 AU in full extent. The minor axis is 42 AU $\pm$ 2
AU peak to peak and 80 AU $\pm$ 5 AU in full extent. Assuming a
position angle of 90$^\circ$, the inclination is then 42$^\circ \pm$
6$^\circ$.

\section{Velocity structure}
\subsection{Spectroastrometry}


Spectroastrometry, determining the line flux spatial mean and standard
deviation in every velocity bin, provides a convenient framework to
model the emission in IRS 48
(e.g. \citealt{pontoppidan08,pontoppidan11b}). Collapsing the spatial
information in this fashion provides several advantages. The biggest
source of uncertainty in the direct imaging is almost certainly the
continuum subtraction. Spectroastrometry, however, is ideally suited
for recovery of weak contrast line signal from strong continuum. While
this is less of a problem in the brightest flux spectral bins of the
$^{12}$CO lines, the contrast in the line wings and weaker transitions
of CO v=2-1 and $^{13}$CO is significantly worse. The more limited
information also makes modeling easier than full 2-D modeling while
retaining the main characteristics of the spatial distribution.


The spectroastrometric signal is the flux-weighted center of the
spectral trace. The weighted mean ($\chi$) was found according to the
formula $\chi=\Sigma x_i F_i / \Sigma F_i$ where $x_i$ is the spatial
position and $F_i$ is the flux at that point. The continuum has a
dilution effect on the observed line signal. The signal that is
actually measured has a contribution from both the continuum and the
line so
$\chi_{Total}(\nu)F_{Total}=\chi_{Total}(\nu)(F_C+F_L)=\chi_C(\nu)F_C+\chi_L(\nu)F_L$,
where subscript $C$ denotes the continuum component and subscript $L$
represents the line component of both the flux ($F$) and the weighted
mean ($\chi$). We assume that the continuum flux is pointlike and thus
$\chi_C=0$. The actual line spatial center is therefore
$\chi_L(\nu)(1+F_C(\nu)/F_L(\nu))$, where $F_C/F_L$ is the continuum
to line flux ratio. The $^{12}$CO lines in IRS 48 are relatively faint
so $F_C/F_L$ is $\sim$9 in the line center, resulting in significant
dilution in the measured signal. $F_C/F_L$ becomes increasingly large
in the line wings as the signal becomes fainter.

To measure the width of the emitting region, we turn to the second
moment or variance ($\sigma^2$), representing the standard deviation
($\sigma$) of the signal. It can be found using the formula $\sigma^2
= \Sigma (x_i - P)^2 F_i / \Sigma F_i$ where $x_i$ and $F_i$ are as
above. $P$ is the central spatial position taken as the peak position
of the continuum emission at 4.7 $\mu$m. The variance is
similarly continuum diluted such that
$\sigma^2_{Total}(\nu)F_{Total}=\sigma_{Total}^2(\nu)(F_C+F_L)=\sigma^2_C(\nu)F_C+\sigma^2_L(\nu)F_L$. However,
the continuum does have a standard deviation due to the telescope
PSF and seeing so $\sigma_C^2 \ne 0$. Here,
$\sigma^2_L=\sigma_{Total}^2(1+F_C(\nu)/F_L(\nu)) - \sigma^2_C
F_C(\nu)/F_L(\nu)$. A further correction to examine the line width
around the line center, so $P=\chi_L$, can be made by subtracting
$\chi_L^2$ so $\sigma^2_{L,P=\chi_L} = \sigma^2_L - \chi_L^2$. The
plots present $\sigma_L$ which can then be compared with the PSF
determined from the width of the continuum signal. Directly
calculating the variance in the line center gives an
emitting region width of 4 AU $^{\rm +10\, AU}_{\rm -4\, AU}$ centered
around the line center. However, this estimate is likely too large as
each spectral pixel covers a range of radii due to the curvature of
the ring. A more detailed and accurate discussion of the line extent
using simple models can be found in Section \ref{kinematics}.

\begin{figure}[h!]
\begin{center}
\includegraphics[angle=90,scale=0.6]{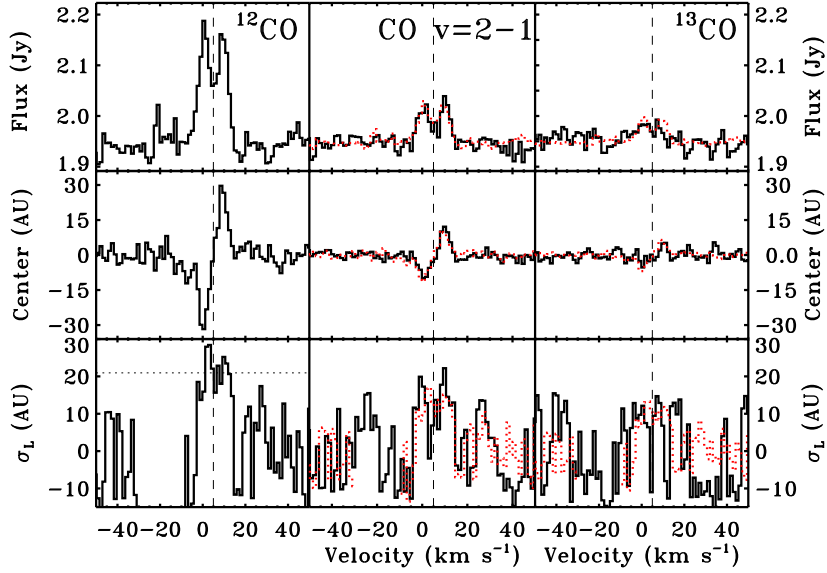} 
\end{center}
\vspace{-1.5cm}
\caption{Comparison of the flux (top), spatial center (middle) and
extent (bottom) of the IRS 48 CO lines for $^{12}$CO (left), CO v=2-1
(middle), and $^{13}$CO (right) at a position angle of 90$^\circ$. The
v=1-0 lines have been scaled to the appropriate flux ratio and
overplotted as dotted red lines in the v=2-1 and $^{13}$CO panels. The
bottom panel shows the standard deviation of the line emission. The
standard deviation of the continuum signal is shown as a dotted
horizontal line in the $^{12}$CO panel. Deconvolving the PSF results
in an emitting region width of 4 AU $^{\rm +10 AU}_{\rm -4 AU}$. The
agreement in spatial postion and line shape of the different
transitions indicates that this emission comes from the same physical
region. \label{fig:isotopes}}
\end{figure}

Figure \ref{fig:isotopes} shows the detected fluxes, spatial centers
and variances of the $^{12}$CO, $^{12}$CO v=2-1 and $^{13}$CO emission
lines. The spectroastrometric signals are calculated over 25 pixels
(129 AU) which includes all the flux, with larger ranges only
increasing the noise. Lines are stacked to enhance the signal. Only
lines without line overlap and central absorption are included. For
$^{12}$CO, this includes P(7), P(8) and P(11). For $^{13}$CO, it
includes R(9), R(11), R(12) and R(13). CO v=2-1 has the most clean
lines as even the low energy rotational levels are free of absorption,
including P(1), P(2), P(3), R(1), R(2), R(5),and R(6). The line shapes
of all the isotopologues are similar but the absolute measured signals
are different due to stronger continuum dilution in the weaker
lines. Multiplying by the correction factors (2.8 for CO v=2-1/CO
v=1-0 and 4.6 for $^{13}$CO/CO v=1-0), results in identical
spectroastrometric signatures within the noise (Figure
\ref{fig:isotopes}). All of the transitions are therefore emitting
from the same physical region.

\begin{figure}[h!]
\begin{center}
\includegraphics[angle=90,scale=0.6]{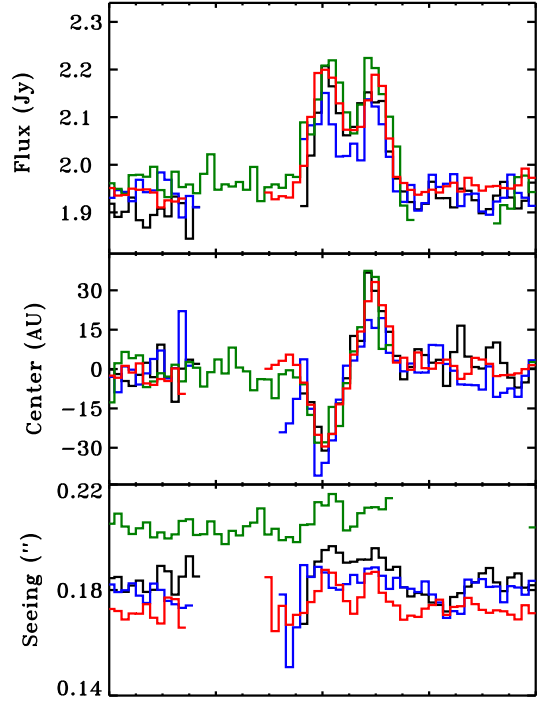} 
\end{center}
\vspace{-1cm}
\caption{Comparison of the flux (top), spatial center ($\chi_{Total}$;
  middle top), and seeing measured from the continuum ($\sigma_{Total}$; bottom) of the IRS 48
  CO lines from different observation dates. Black and blue are 2 Aug,
  2008 at PA of 90 and 270 respectively (spatial center for
  270$^\circ$ is multiplied by -1). Red is 5 September, 2007 and green
  is 3 May 2008. \label{fig:variability}}
\end{figure}

Figure \ref{fig:variability} shows the spectroastrometric signatures
arising from the different epochs. Observation in three different
epochs 3-9 months apart allows us to search for variability. While the
continuum width varied between the different nights due to AO
performance, the CO structure appears stable over the year timescale.

\subsection{Kinematic modeling}
\label{kinematics}

A simple Keplerian disk model is fitted to the data to examine the
physical structure of the CO (following \citealt{pontoppidan08}). The
model is gridded in the disk frame with rings of gas at the Keplerian
velocity. Each ring is assigned a temperature according to an overall
disk temperature profile. The surface density is assumed to be
constant with a total column density, $N_{CO}$. The local line width
is set to 3 km s$^{-1}$, the instrumental spectral resolution. The
flux at each grid point is determined for the different rotational
lines. The grid is then transformed into the observer's Cartesian
coordinates by twisting to the correct position angle and
inclination. The emission is so extended that the 0\farcs2 slit does
not capture all the light, so the data is corrected for the truncation
accounting for the AO-corrected PSF of 0\farcs18 based on the measured
continuum width (see Fig. \ref{fig:variability}). The emission in
the PA=0$^\circ$ position is strongly limited by this parameter.

\begin{figure*}[h!]
\begin{minipage}{0.5\linewidth}
\begin{center}
\includegraphics[angle=90,scale=0.31]{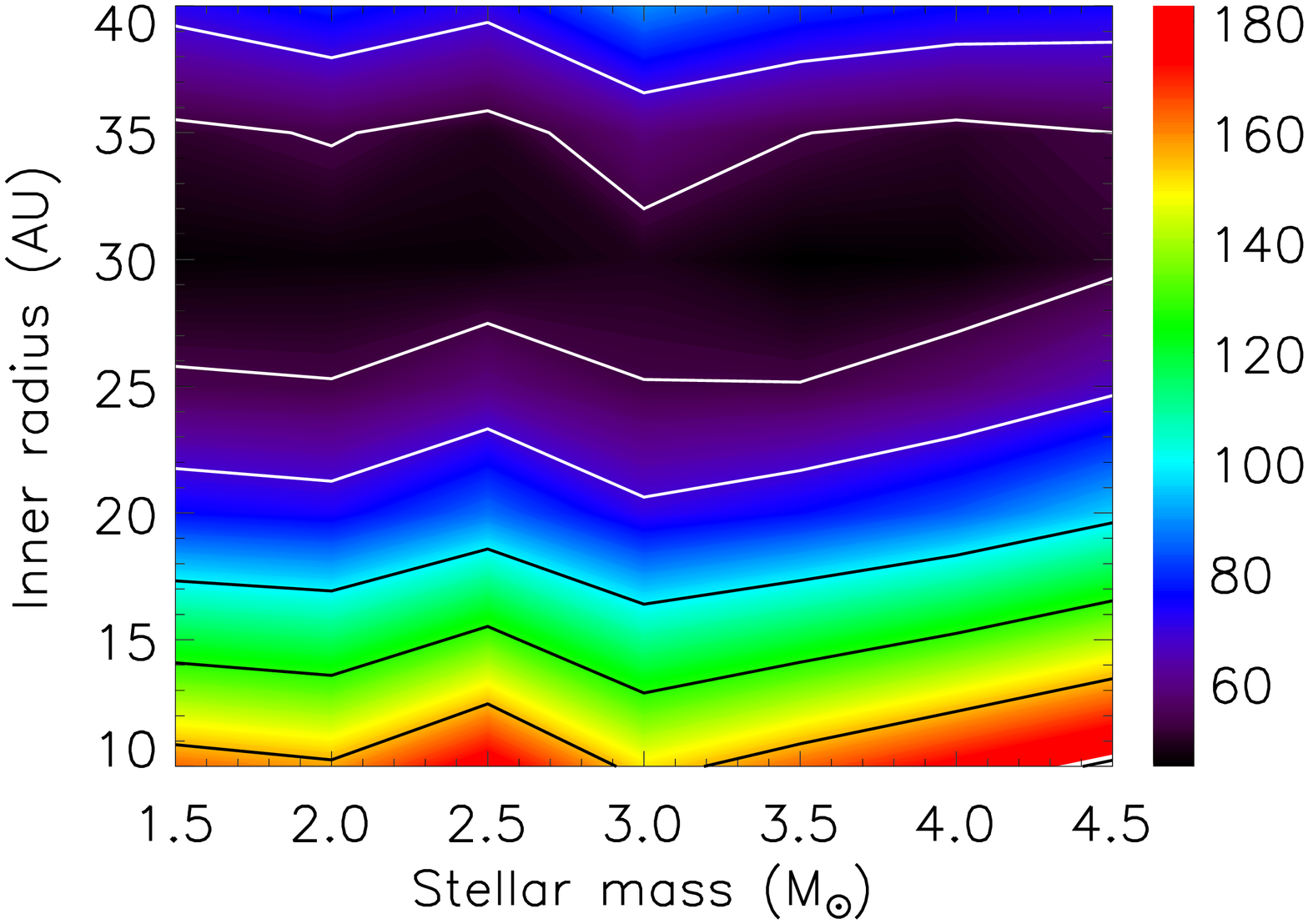} 
\end{center}
\end{minipage}
\begin{minipage}{0.5\linewidth}
\includegraphics[angle=90,scale=0.3]{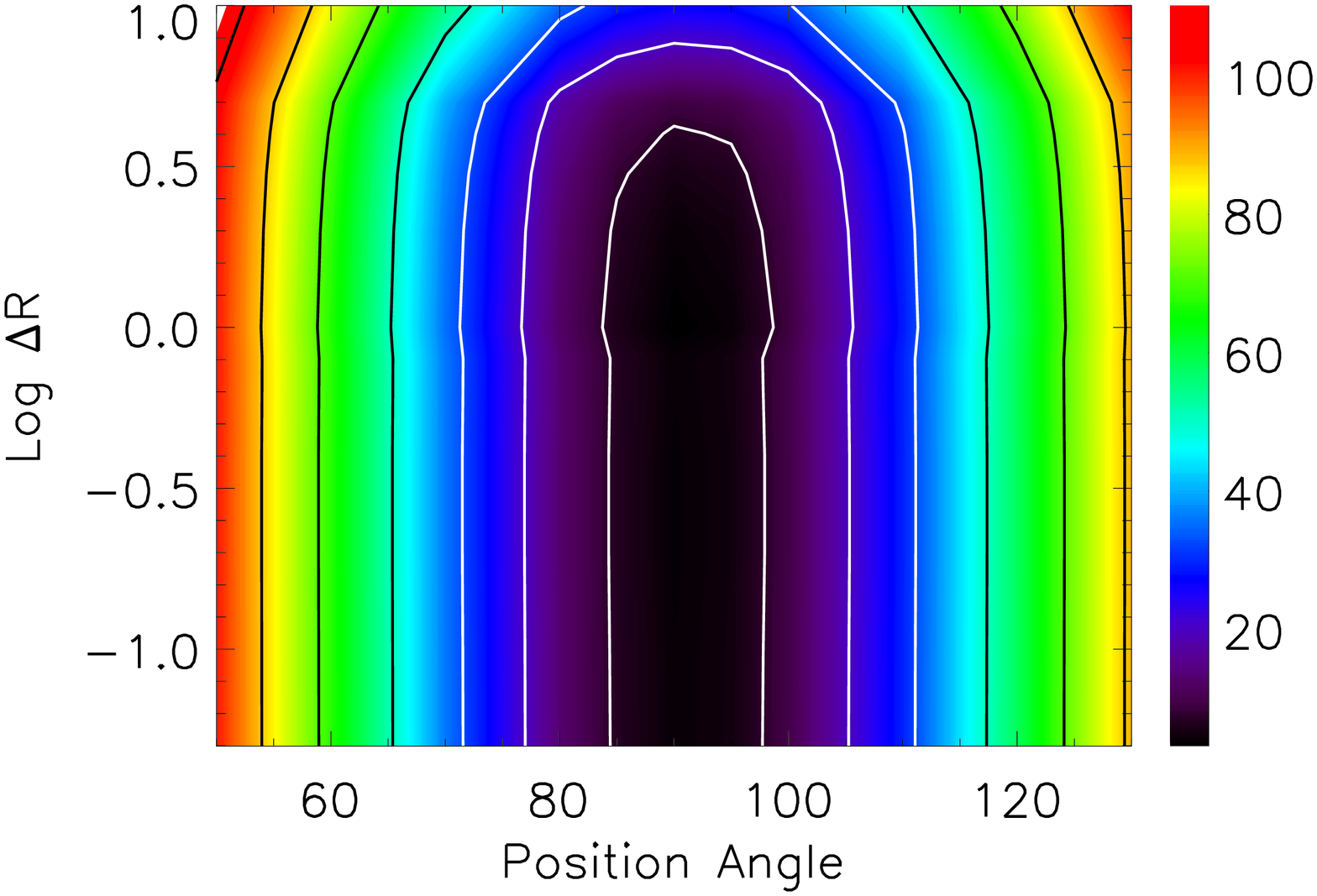} 
\end{minipage}

\caption{(left) $\chi^2$ plot of $R_{0}$ at each $M$, choosing the
inclination and position angle which produce the best fits. The inclinations
range from 30$^\circ$ to 50$^\circ$ with the higher masses tending to
better fits at smaller inclinations. $R_0$ is therefore fairly robust
across a range of stellar masses. (right) The effects of varying
emission region widths in a uniform temperature ring with constant
density. The emission is clearly dominated by a small region close to
$R_0$. \label{fig:chi2}}
\end{figure*}

A grid of models varying the stellar mass, inclination, inner radius
($R_0$) and position angle were run to examine the interdependence of
these parameters. The two most interesting parameters are the inner
radius and the stellar mass. In order to robustly determine the inner
radius regardless of mass, the minimum $\chi^2$ value for each mass
was determined. Figure \ref{fig:chi2} plots mass against $R_0$,
assuming the inclination associated with the minimum value. The hole
size ($R_0$) minimum is 30 AU throughout the mass range with optimal
inclinations of 50$^\circ$ at the low mass end and 30$^\circ$ at the
high mass end (Figure \ref{fig:chi2}).

\begin{figure}[h!]
\includegraphics[angle=90,scale=0.31]{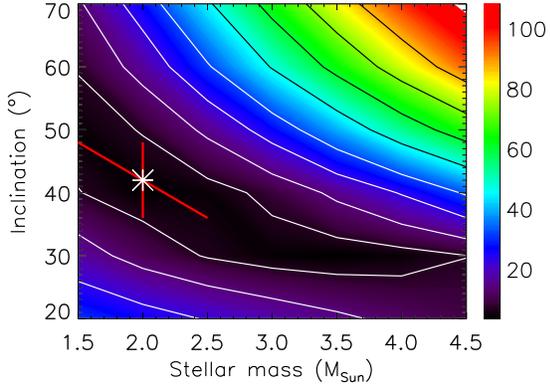} 
\caption{$\chi^2$ plot of stellar mass vs inclination. The white star
marks where the measured inclination of 42$^\circ$ is at the $\chi^2$
minimum.The red vertical error bars show the error on the inclination
while the diagonal line traces the lowest $\chi^2$ values within the
inclination errors. \label{fig:mincl}}
\end{figure}

Stellar mass is degenerate with inclination in these
models. Accurately determining the inclination is therefore the best
method of determining the stellar mass. The inclination from
\citet{geers07} is 48$^\circ\pm8^\circ$, which becomes incompatible
with masses higher than 3 M$_\odot$. The CRIRES 2D images give a
slightly more face-on inclination measurement of
42$^\circ$$\pm$6$^\circ$ (see \S \ref{analysis}). Figure
\ref{fig:mincl} shows the $\chi^2$ values for stellar mass versus
inclination with the CRIRES inclination overlaid. The stellar mass of
2 M$_\odot$ $\pm$ 0.5 M$_\odot$ agrees with the determination
from the optical spectra and is incompatible with a 1 Myr A0 star.

As the surface density is held constant, the temperature profile is
also a proxy for the distribution of gas. We test two different
distributions, a power-law and a uniform temperature ring. The power
law distribution is of the form, $T=T_0(R/R_0)^{-q}$, where $R_0$ is
taken to be the inner most edge of the gas and the power-law declines
from there until the outer edge of the disk, set at 100 AU. $T_0$ is
set at 260 K in agreement with the rotation diagram results from
section \ref{rot}. A steep power law index ($q$$>$0.4 for best
results) is required in order to show the double peaked flux
profile. The uniform temperature ring is centered at $R_0$ and $T_0$
but has width $\Delta R$. The effects of varying emission widths in a
uniform temperature ring with constant density are shown in Figure
\ref{fig:chi2}. The width of the emitting region is directly related
to the variance, $\sigma_L^2$. The best fits result from the narrowest
emitting regions, limited only by the grid cell size. Overall there
was little difference between the two distributions as the good fits
from both concentrated the emission close to $R_0$ with a width of a
few AU or less. Physically this narrow region is likely the dust hole
cavity wall.

\begin{figure*}[h!]
\begin{center}
\includegraphics[angle=90,scale=0.8]{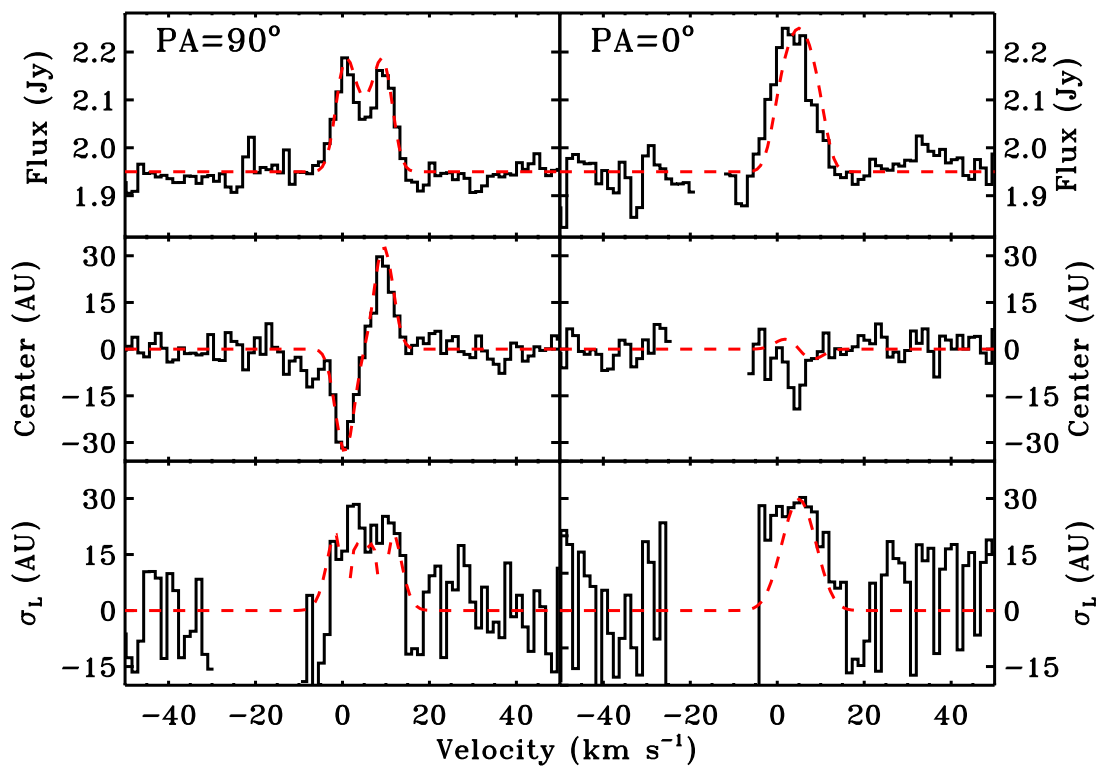} 
\end{center}
\vspace{-3cm}
\caption{Comparison of the flux (top), spatial center (middle) and
extent (bottom) of the IRS 48 CO lines. The left column is the
spectroastrometry at position angles of 90$^\circ$ and 270$^\circ$
while the right column is 0$^\circ$ and 180$^\circ$. The overplotted
red line is a simple model of a ring of gas in Keplerian rotation at
30 AU around a 2 M$_\odot$ star with an inclination of 42$^\circ$. The
temperature of the disk outside this radius is taken to have a
power-law distribution with an index of -0.6.\label{fig:model}}
\end{figure*}

In summary, the best fit to the data was obtained using the derived
stellar mass of 2 M$_\odot$ (\S \ref{stellarparams}) and an
inclination of 42$^\circ$ from the 2-D CRIRES data (Figure
\ref{fig:model}). The hole size is 30 AU with the disk at a position
angle of 95$^\circ$, in good agreement with the mid-IR image. The
model fit of the PA=0$^\circ$ observations is not as good due to the
asymmetries in the data (i.e. Figure \ref{fig:p5}). The structure
appears more complex than a simple uniform ring in that direction.

\section{Excitation}

\subsection{Rotational temperatures}
\label{rot}

Rotation diagrams, which use the relative strengths of the rotation
lines to understand the gas temperature and column density, were
constructed for both the emission and absorption
components. Interpretation of rotation diagrams is commonly based on
the Boltzmann distribution and the assumption of LTE level
populations, so 
\begin{equation}
\frac{n_J}{g_J} = \frac{n_{CO}}{Q_{Rot}}e^{(E_J/kT)},
\end{equation}
 where n
is the number, $Q_{Rot}$ is the partition function, and $E_J$ is the
energy of level $J$. We plot ln $(F_J/gA\nu)$ against $E_{\rm
Upper}/k$ where $F_J$ is the line flux, $g$ is the degeneracy
(2$J_{Up}$+1), $A$ is the Einstein A coefficient, $\tilde{\nu}$ is
wavenumber in cm$^{-1}$ and $E_{\rm Upper}$ is the upper state
energy. In the optically thin case, a linear fit will match the data
well. The slope ($\alpha$) is related to the temperature, $T_{Rot}$
such that $T_{Rot}=-1/\alpha$. The intercept, $y$, is related to the
CO column density, $N_{CO}$, such that
\begin{equation}
N_{CO}=Q_{Rot}e^y\frac{16\pi d^2}{h c A_{emit}}
\end{equation}
where $d$ is the distance and $A_{emit}$ is emitting area.

\subsubsection{Absorption lines}

The absorption lines are strongest in the lowest energy rotational
lines and absorption features can be seen in the less abundant
isotopes, $^{13}$CO, C$^{18}$O and C$^{17}$O. The absorption lines are
spectrally unresolved at 3 km s$^{-1}$ and have a central v$_{\rm
LSR}$ of 3.8 km s$^{-1}$ in agreement with the $\rho$ Ophiuchi
region (e.g. \citealt{difrancesco04, maruta10}).  The derived gas
excitation temperatures from the different isotopologues are $\sim$25 K (see
bottom half of Table \ref{table:rot}). The $^{12}$CO absorption is also highly
optically thick. These cold temperatures are consistent with an origin
in foreground cloud material. The absorption lines are therefore
probably not related to the disk and are excluded from further
analysis.

\subsubsection{Emission lines}

Quantifying the emission line fluxes is complicated by the
contaminating absorption lines at low J. A clean template of the emission line
profile is made by stacking the higher J lines which are free from
absorption. This template proves to be a good match for the CO v=2-1
and $^{13}$CO emission lines as well as the CO v=1-0.  The profile is
then fitted via chi-squared minimization on top of a measured local
continuum to derive each line flux. This method provides more
consistent measurements than integrating the flux directly which is
strongly affected by noise. The low $J$ $^{12}$CO lines are
contaminated at low velocities by absorption, so a fit was made to the
line wings.

\begin{figure}[h!]
\begin{minipage}{0.5\linewidth}
\includegraphics[angle=90,scale=0.35]{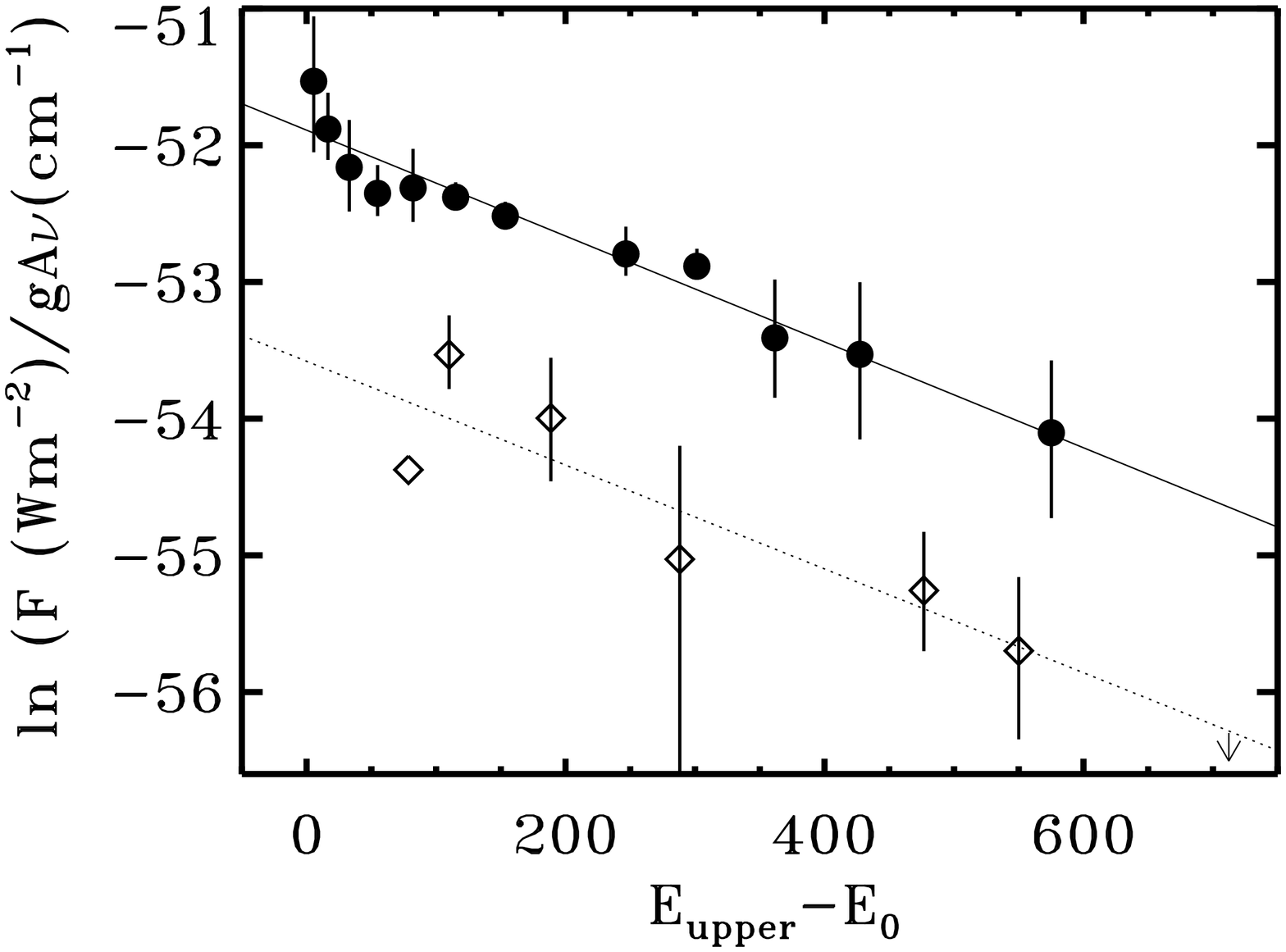} 
\end{minipage}
\begin{minipage}{0.5\linewidth}
\includegraphics[angle=90,scale=0.35]{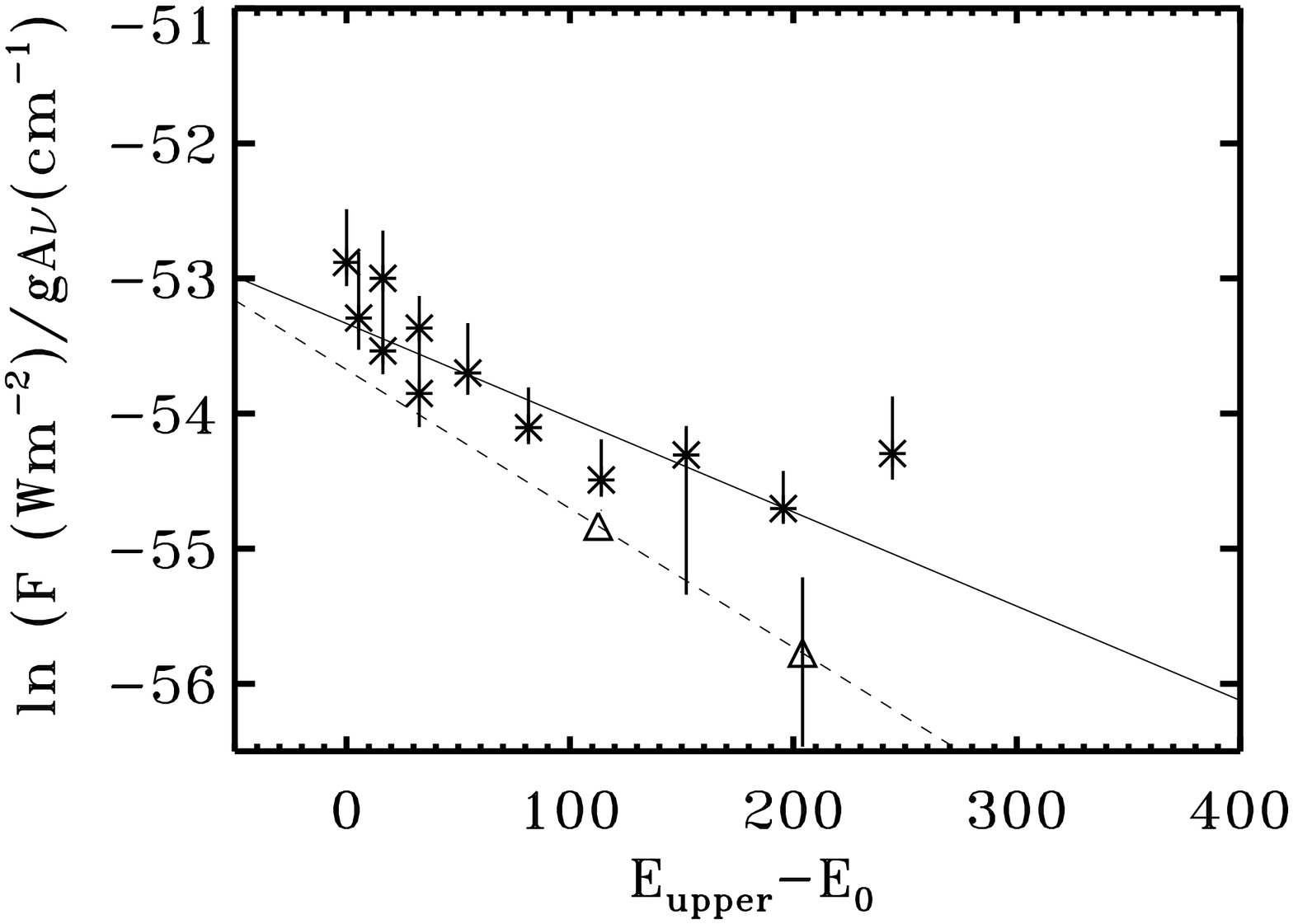} 
\end{minipage}
\caption{Rotation diagram of the CO emission lines. Filled circles are
$^{12}$CO, diamonds are $^{13}$CO (dotted fit), stars are $^{12}$CO
v=2-1, and triangles are $^{12}$CO v=3-2 (dashed fit). The energies
refer to those of the rotational levels relative to J=0 in the
respective vibrational state.\label{fig:emrot}}
\end{figure}

The rotation diagram (Figure \ref{fig:emrot}) for $^{12}$CO v=1-0,
$^{12}$CO v=2-1 and $^{13}$CO relative to the energy of J=0 in each
vibrational level indicates that the different transitions have
similar properties. The distributions all appear linear, in agreement
with optically thin gas at a single temperature. In this case, the
derived rotation temperatures are $\sim$250 K (Table
\ref{table:rot}). Small changes in the derived slope
(i.e. temperature) can result in large changes in derived
population. All the slopes are in the range 100-260 K, but the better
constrained species are at the higher end. The total mass ($N_{CO}\,
{\rm x}\, A_{emit}$) is well constrained, but the column density is
not due to the lack of constraints on the emitting area. Based on the
kinematics (\S \ref{kinematics}), we estimate that the area is a ring
at 30 AU with a width of 0.5 AU and use this area to calculate the
column densities in Table \ref{table:rot}. A gas temperature of 250 K
is slightly high compared to the likely dust temperature of $\sim$150
K at 30 AU based on the low luminosity of the star. However, the gas
temperature may be much higher than the dust temperature due to strong
UV radiation. The CO v=2-1 lines are too strong to be purely thermally
excited with a vibrational temperature much higher than the rotational
temperature. We attribute the excess excitation to UV fluorescent
excitation which is discussed in Section \ref{vib}. The low CO
v=1-0 lines may be slightly optically thick but the errors are
large due to the central absorption lines and a linear fit is well
within the errors. However, the optically thin fits produce a low
isotopic ratio $^{12}$CO/$^{13}$CO of 15, challenging this assumption.

\begin{figure}[h!]
\begin{center}
\includegraphics[angle=90,scale=0.4]{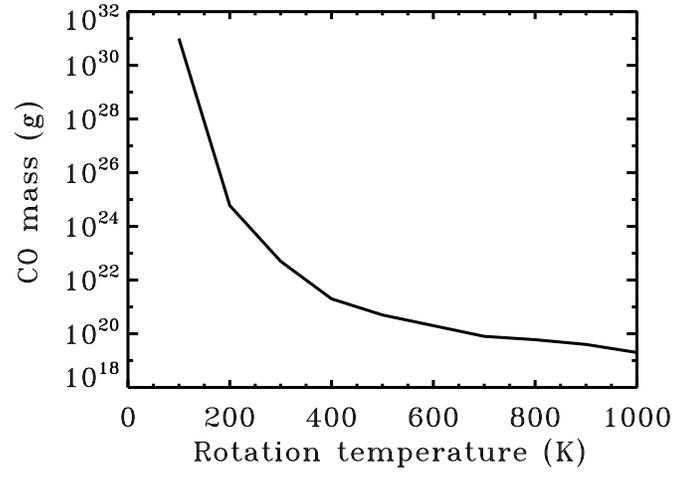} 
\end{center}
\caption{Upper limits on CO mass within the 30 AU hole assuming
different gas temperatures. \label{fig:coinholelimit}}
\end{figure}

In order to determine a limit on the amount of CO within the hole,
flux limits were determined from the 2-D images. In general, the flux
levels were similar to residuals from the continuum subtraction in
regions without CO emission and were on the order of 10\% of the peak
flux. Assuming optically thin gas, rotation diagrams for gas at
different temperatures were compared with the flux limits to determine
the maximum CO gas mass that could be within the hole without
detection (Figure \ref{fig:coinholelimit}). The amount is strongly
dependent on temperature with much lower limits at high temperatures
and weaker constraints at low temperatures. In general, the inner
region should be hotter than the gas further out. The mass limit is
$\sim$10$^{20}$ g for temperatures above 500 K. This translates to a
gas density of $\sim$10$^3$ cm$^{-3}$, assuming a uniform distribution
within the 30 AU hole and a disk height resulting from hydrostatic
equilibrium.

\subsection{UV fluorescent excitation}
\label{vib}

The distance of the gas from the star, resulting in cooler
temperatures, and the presence of high energy CO v=2-1 and v=3-2 lines
suggest that a non-equilibrium process such as UV fluorescence is
likely responsible for their line emission. The measured vibrational
temperature is 2200 K, greatly in excess of the thermal rotational
temperature of 260 K. A strong UV field is to be expected from an A
type star such as IRS 48 and the strong PAH emission from within the
hole requires its presence.

A simple UV fluorescence model was run to examine the likelihood of
the stellar UV field producing the observed vibrationally excited
emission at 30 AU. The model follows \citet{brittain07} but expands
the vibrational levels covered to 35 $X^1\Sigma^+$ and 25 $A^1\Pi$
levels. The model assumes statistical equilibrium so each level
maintains a steady population. We then considered excitation from
$X^1\Sigma^+$ to $A^1\Pi$ ($g_{A-X}$) balanced by spontaneous emission
($A_{A-X}$) back to the ground electronic state. Decay out of the
excited electronic state populates a range of vibrational levels in
the ground state, including vibrational levels which are too energetic
to be thermally populated. We also include spontaneous relaxation
between vibrational levels in the ground electronic state ($A_{X-X}$),
which occurs primarily between levels with $\Delta {\rm v} = 1$. The
transitions covered between the ground state $X^1\Sigma^+$ electronic
state and the $A^1\Pi$ excited electronic state occur at ultraviolet
wavelengths 1270 to 4000 \AA, but the most populous levels are mainly
driven in the range 1270 to 1700 \AA. The model also includes
collisional excitation, assumed to be thermalized at $T_0$ = 260 K.

In general, the model output depends purely on the input UV
spectrum. While accretion shocks produce much of the UV emission in
lower mass T Tauri stars, the similarity between the photospheric and
shock temperatures in Herbig Ae stars results in the photospheric
emission dominating \citep{valenti00}. Based on our derived stellar
properties, we constructed a 9000 K blackbody and compared this to an
HST STIS UV spectrum of AB Aur (from
StarCAT\footnote{http://casa.colorado.edu/$\sim$ ayres/StarCAT/}). AB Aur is
a good proxy for IRS 48 with a spectral type of A0 and age of 4
Myr. While the spectrum contains many lines, the overall flux and
shape is a good match to the 9000 K blackbody, once scaled to the
appropriate luminosity. The calculations used the observed
spectrum in the available range (1140-1729 \AA), which accounts for
99\% of the transitions in this source, and the blackbody at longer
wavelengths.

The observed fractional population in each vibrational level was
calculated using the rotational diagrams (Figure \ref{fig:emrot}). The
energies of each rotational state were normalized to the zero point of
each vibrational level. The intercept was used to derive the
population in each level, which was then divided by the total
population. The ground state v=0 is by far the most populated. The low
signal-to-noise in many of the faint lines introduces a lot of scatter
in the rotational diagram. Given the similarity in line profiles and
spectroastrometry, the gas is likely all close in temperature so we
fix the rotational temperatures at 260 K and then derive the
populations.

\begin{figure}[h!]
\begin{center}
\includegraphics[angle=90,scale=0.4]{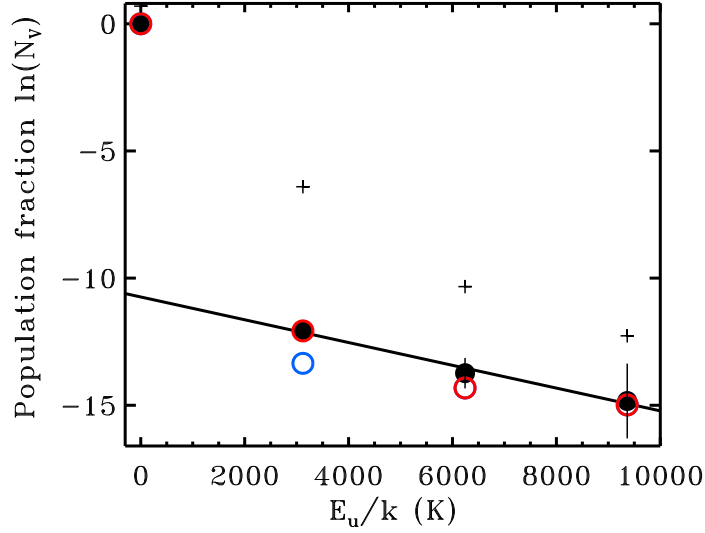} 
\end{center}
\caption{Vibrational diagram for IRS 48. The black points are the
observed fractional population in each vibrational level. The linear
fit indicates a vibrational temperature of 2200 K, greatly in excess
of the rotation temperature of 260 K. The red circles are the expected
vibrational fractions from a UV fluorescence model of IRS 48,
including the thermal population. The blue circles, underlying the red
for v$>$1, are the vibrational component in each level before the
collisionally excited population is added. The black crosses represent
the expected excitation from our 1 Myr old, high mass spectral
parameters with a luminosity of 200 L$_\odot$ at 30 AU. While the
vibrationally excited emission increases by about an order of
magnitude, the bulk of the increase is from the expected increase in
collisional excitation with the increased temperature at 30
AU. \label{fig:vibplot}}
\end{figure}

Figure \ref{fig:vibplot} shows the vibrational diagram with the
overlaid model fit. A linear fit to the data (v $>$ 0) gives a
vibration temperature of 2200 K, significantly higher than the
rotational temperature. However, the thermal component to the v'=1
level is significant, actually causing the vibrational temperature to
be underestimated. The model of the effect of the UV field at 30 AU,
including the thermal contribution from gas at 260 K under the
assumption of LTE excitation, is shown in red while the purely
vibrational excitation is shown in blue. For v=2 and 3, the blue
points lie completely under the red as there is virtually no
collisional excitation to these levels. The vibrational temperature of
the model is $\sim$5000 K. The model fit is in excellent agreement
with the data, indicating that the expected UV field from IRS 48 for
$M_*$ = 2 M$_\odot$ and $L_*$ = 14.3 L$_\odot$ can produce the observed
energy distribution.

We also investigate the expected excitation from the younger, higher
mass star implied by the evolutionary age of $\rho$ Oph (\S
\ref{stellarparams}). The luminosity is set at 200 L$_\odot$ and the
distance is kept at the 120 pc distance to $\rho$ Oph. The black
crosses in Figure \ref{fig:vibplot} represent the expected excitation
from such a star at 30 AU. While the vibrationally excited emission
increases by about an order of magnitude, the collisional excitation
also increases due to a higher temperature of 500 K at 30 AU. It
appears unlikely that IRS 48 is a younger, more luminous, but highly
extincted star. One remaining potential source of error in the
luminosity would be an incorrect distance, implying that IRS 48 is not
actually in the $\rho$ Oph molecular cloud. The vibrational excitation
does not provide a constraint on this scenario as a larger distance
simply corresponds to a large hole size and the vibrational diagram is
unchanged.

\section{Discussion}
\label{discussion}

Based on our CRIRES data, the CO emission around IRS 48 comes from a
ring at 30 AU in Keplerian rotation around the central star. We
propose that the CO gas is thermally and UV excited in the dust wall
at the hole edge. The radius of the gas emission and the dust hole
edge seen in the \citet{geers07} 18.7 $\mu$m dust image are similar,
indicating that the CO and dust rim are co-spatial. Our Keplerian
model is consistent with a ring of emission with little to no flux
required from outside the hole wall region, as seen in the required
steep power law index. The majority of the stellar UV light must reach
the hole wall in order to excite the CO v=2-1 and v=3-2 emission seen
along with the v=1-0 emission, given the good match between the
fluorescence model and data.  While little dust shielding is likely
present inside the hole, the PAH molecules could absorb some of the UV
field.  Although without knowing either the abundance or distribution
of the PAHs, it is difficult to estimate the magnitude of the
absorption. The hole wall may be vertically puffed up so that a
fraction of the stellar UV light reaches the rim unattenuated.

Despite the lack of CO interior to 30 AU, some material is present in
the inner disk. The strong, centrally-peaked PAH emission seen by
\citet{geers07} indicates that small particles remain within the dust
hole in IRS 48. Generally, PAHs are small enough that they are expected to be
dynamically coupled to the gas. The presence of PAHs also promotes the
formation of H$_2$, CO and other small molecules even in the absence
of small grains \citep{jonkheid06}. Other
signs of material inside the hole are the H$\alpha$ lines in both our
2008 and the \citet{geers07} optical spectra and the Pf$\beta$ line
present in all the CRIRES data - all probable signs of ongoing
accretion. We know little about the distribution of gas within the
hole from these signatures although material is likely still reaching
the star.

Any overall picture of the IRS 48 system must therefore explain both
the truncation of the dust grains and CO at 30 AU as well as the
destruction of CO, but not PAHs and atomic hydrogen, within the
hole. The strong UV field within the hole may be the cause of the
different survival rates.  Larger PAH molecules have a better chance
of surviving in high UV environments than smaller ones. According to
\citet{visser07}, PAHs of sizes 50 carbon atoms and smaller are
destroyed at the $\tau$=1 surface out to 100 AU around a Herbig Ae/Be
star. Given the likely low opacity of the disk interior only PAHs with
100 carbon atoms or greater would be able to populate the hole outside of
about 5 AU, sufficient to explain the mid-IR VISIR images. On the
other hand, CO has a photodissociation lifetime of $<$ 1 year inside
the hole without shielding. Preliminary modeling indicates that gas
densities below 10$^{5}$ cm$^{-3}$ are inadequate for rapid
reformation of CO and prevent self-shielding (Bruderer,
priv. comm.). Our limits on the CO gas density within the hole lie
well below this value for temperatures greater than 500 K. However, the
CRIRES data can give little constraint on any low temperature gas. Gas
with temperatures of 300 K or less could be present at densities of
10$^5$ cm$^{-3}$. This regime of PAH emission but no CO emission
places stringent limits on the amount of gas present in the hole.

The sharp truncation of CO and dust points towards either physical
truncation by a companion or photoevaporation of the CO. While the UV
field around IRS 48 is strong and the accretion signatures are weak,
it is not clear that photoevaporative inside-out clearing would leave
the material inside the hole in this state
(e.g. \citealt{alexander06}, \citealt{owen11}). Gas and dust inside
the hole in the photoevaporative clearing scenario is usually residual
material still draining onto the star and is not being
replenished. This system would thus have to be in a specific time
period where the gap has open, the density of the interior material
has dropped below the CO photodissociation threshhold but has not yet
drained completely.  A companion, on the other hand, would naturally
reduce the flow of material inwards and create a drop in density of
both gas and dust. A final possibility, although probably the least
likely, is that no material is flowing through the hole and the PAHs
are formed in situ via the collision of large rocky bodies.

The CO hole in the Oph IRS 48 disk is the largest gas ring discovered
to date for any pre-main sequence star. The hole is more than twice as
large as the 11 AU holes around the Herbig Ae/Be stars, HD 141569 and
HD 97084, which have similar spectral types to IRS 48. SR 21, a G-type
star, with a ring of emission at 7 AU is the lowest mass source with a
known large near-IR CO hole \citep{pontoppidan08}. Vibrational
emission of similar extent to the v=1-0 lines in all cases indicates
that the UV field may be the driving force in maintaining the presence
of the CO holes via photodissociation. Large near-IR CO holes of $>$1
AU have not yet been found in K- or M-type T Tauri transition disks,
despite the presence of substantial dust holes. In fact, many
transition disks contain CO close to the star at radii significantly
smaller than the dust hole size \citep{pontoppidan08,salyk11}. The UV
energy from T Tauri stars is significantly smaller than from higher
mass stars with a large portion generated in accretion shocks. If
transition disks have even smaller accretion rates than classical T
Tauri stars as has been suggested \citep{najita07}, T Tauri stars are
unlikely to have both an absence of absorbing inner dust and a strong
UV field, hampering the creation of large photodissociated CO
holes. Although CO fundamental emission is detected from the inner
dust wall of IRS 48, any non-detection of CO emission from T Tauri
disks with holes (e.g. CoKu Tau/4) may be caused by difficulty
exciting the 4.7 $\mu$m CO lines at large radii due to both the weaker
UV field and lower temperatures, rather than a lack of gas at the
inner rim of the dust disk. Further work to better understand the
excitation of gas in holes is needed to determine if the creation of
such large CO holes is a unique phenomenon of intermediate mass stars.

We are also left with an evolutionary puzzle regarding IRS 48. The
CRIRES data, both from kinematics and excitation, are much more
consistent with a lower mass star of around 2 M$_\odot$ and a
luminosity of $\sim$15 L$_\odot$, as derived from the optical
spectroscopy. The data are inconsistent with a prototypical 1 Myr A0
star with a mass of 4.5 M$_\odot$ and luminosity of 200
L$_\odot$. However, the resulting age from evolutionary tracks of
$\sim$15 Myr is much older than the $\sim$1 Myr age of Ophiuchus
\citep{luhman99}, with the star being underluminous by more than an
order of magnitude. Such an old PMS star might be expected to have a
debris disk of reprocessed dust and no gas, but this is clearly not
the case for IRS 48. IRS 48 appears to be a member of Ophiuchus with
proper motions in agreement with the cloud (E. Mamajek, priv. comm.),
a position close to the core, emission lines centered at the cloud
velocity and an extinction consistent with a location within or behind
the cloud. \citet{baraffe09} suggest that episodic cold accretion
leads to different contraction timescales, and thus some stars may
appear much older from evolutionary models than their true age. While IRS 48 is more massive
than the stars studied by \citet{baraffe09}, similar mechanisms may
apply even at higher mass and help explain the apparent age
discrepancy for IRS 48. If the stellar parameters, cloud membership
and PMS evolutionary tracks are all correct, Oph IRS 48 would be an
extremely old member of Ophiuchus, possibly left over from a previous
epoch of star formation.

\section{Conclusions}

We have used the combination of high spatial and spectral resolution
of CRIRES to observe spatially extended CO emission from Oph IRS
48. IRS 48 was known to be a transitional disk from the imaged 18
$\mu$m dust ring but the centrally peaked PAHs lead to an expectation
of interior gas \citep{geers07}. Despite the presence of PAHs within the
dust hole, the CO emission peaks at 30 AU, co-spatial with the dust
hole edge, and shows no emission interior to the dust hole. The
spatial distribution of the vibrationally excited gas and
isotopologues is the same, indicating a common origin. The CO geometry
is in agreement with the \citet{geers07} dust image with an east-west
major axis, brighter emission from the north side compared to the
south, and a derived inclination of 42$^\circ\pm$6$^\circ$. The gas is
thermally excited to a temperature of 260 K, but is also strongly
fluoresced by the central star's UV field, showing vibrational
excitation temperatures of $\sim$5000 K. The new spectral type of
A0$^{+0}_{-4}$ resolves the discrepancy of strong PAHs around what was
thought to be an M0 star \citep{geers07} and explains the
vibrationally excited emission. Our kinematic modeling is consistent,
both spectrally and spatially, with a ring of gas at 30 AU in
Keplerian rotation with a width of only a few AU or
less. All the evidence points towards the gas being UV excited in the
outer rim of the hole. A sudden drop in density to $<$10$^5$
cm$^{-3}$, as might be caused by a companion, could cause the CO to
photodissociate on short timescales but maintain the larger PAH
population. IRS 48 is therefore an excellent target for planet finding
follow-up.




\acknowledgments {The authors wish to thank Vincent Geers for
providing the reduced data from his 2007 paper and Eric Mamajek for
discussion of the cloud membership of IRS 48. J.M. Brown acknowledges
the Smithsonian Astrophysical Observatory for support from a SMA
fellowship. Astrochemistry at Leiden is supported by a Spinoza grant
from the Netherlands Organization for Scientific Research (NWO) and by
the Netherlands Research School for Astronomy (NOVA) grants. }

\bibliographystyle{apj}

\newpage

\newpage

\newpage

\newpage

\newpage

\newpage

\newpage

\newpage

\newpage

\newpage

\newpage

\newpage

\newpage

\newpage

\begin{deluxetable}{lccccl}
\tablecolumns{6}
\tablewidth{0pt}
\tabletypesize{\normalsize}
\tablecaption{\label{table:obs}Summary of CRIRES observations}
\tablehead{\colhead{Date} & \colhead{Wavelength} & \colhead{Position} &
\colhead{Number of } & \colhead{Standard} \\
\colhead{} & \colhead{(nm)} & \colhead{Angle} &
\colhead{Nod Cycles} & \colhead{Star}}
\startdata
2007 Sep 5 & 4730 & 90 & 1 &  HR 5812\\
2008 May 3 & 4730, 4833 & 90 & 2 &  HR 6084\\
2008 Aug 2 & 4710 & 0 & 3 &  HR 6084\\
2008 Aug 2 & 4730 & 0,90,180,270 & 3  & HR 6084\\
2008 Aug 5 & 4868 & 90 & 3.5 & HR 6175 
\enddata
\end{deluxetable}

\newpage

\begin{deluxetable}{lcc}
\tablecolumns{3}
\tablewidth{0pt} \tabletypesize{\normalsize}
\tablecaption{\label{table:model}Model parameters}
\tablehead{\colhead{Parameter} & \colhead{Best fit model} & \colhead{Range}} 
\startdata 
Hole size ($R_0$) & 30 AU & 25-35 AU \\
Stellar mass & 2 M$_\odot$ & 1.5-2.5 M$_\odot$ \\
Inclination & 42$^\circ$ & 36-48$^\circ$ \\
Position angle & 95$^\circ$ & 85$^\circ$-100$^\circ$ \\
q & 0.6 & $>$0.4 \\
$\Delta R$ & 1 AU & $<$ 3 AU 
\enddata 
\end{deluxetable}

\newpage

\begin{deluxetable}{lccc}
\tablecolumns{4}
\tablewidth{0pt} \tabletypesize{\normalsize}
\tablecaption{\label{table:rot}Summary of rotation diagrams$^1$}
\tablehead{\colhead{Molecule} & \colhead{Rotational} & \colhead{CO Mass}
& \colhead{CO Column Density}\\ \colhead{} & \colhead{Temperature (K)} & \colhead{(g)}
& \colhead{(cm$^{-2}$)}} 
\startdata 
$^{12}$CO v=1-0 & 256 & 7.2 10$^{23}$ & 7.2 10$^{17}$ \\ 
$^{13}$CO v=1-0 & 267 & 4.6 10$^{22}$ & 4.6 10$^{16}$ \\ 
$^{12}$CO v=2-1 & 204 & -- & -- \\ 
\hline 
$^{12}$CO v=1-0 & 26 & -- & -- \\
$^{13}$CO v=1-0 & 17 & -- & -- \\ 
C$^{18}$O v=1-0 & 30 & -- & --
\enddata 
\tablenotetext{1}{Emission rotation diagram parameters are
above the horizontal divide and absorption parameters are below.}
\end{deluxetable}

\clearpage




\end{document}